\documentclass[amsfonts,twocolumn]{revtex4-1}
\usepackage[applemac]{inputenc}
\usepackage{graphicx,hyperref}

\def\vec#1{{\bf #1}}

\graphicspath{{figures/}}

\begin{document}

\title{Thermodynamics of anisotropic triangular magnets with
ferro- and antiferromagnetic exchange}

\author{Burkhard Schmidt and Peter Thalmeier}
\affiliation{Max-Planck-Institut f{\"u}r Chemische Physik fester
Stoffe, 01187 Dresden, Germany}

\date{\today}

\begin{abstract}
We investigate thermodynamic properties like specific heat $c_{V}$ and
susceptibility $\chi$ in anisotropic $J_1$-$J_2$ triangular quantum
spin systems ($S=1/2$).  As a universal tool we apply the finite
temperature Lanczos method (FTLM) based on exact diagonalization of
finite clusters with periodic boundary conditions.  We use clusters up
to $N=28$ sites where the thermodynamic limit behavior is already
stably reproduced.  As a reference we also present the full
diagonalization of a small eight-site cluster.  After introducing
model and method we discuss our main results on $c_V(T)$ and
$\chi(T)$.  We show the variation of peak position and peak height of
these quantities as function of control parameter $J_2/J_1$.  We
demonstrate that maximum peak positions and heights in N\'eel phase
and spiral phases are strongly asymmetric, much more than in the
square lattice $J_1$-$J_2$ model.  Our results also suggest a tendency
to a second side maximum or shoulder formation at lower temperature
for certain ranges of the control parameter.  We finally explicitly
determine the exchange model of the prominent triangular magnets
Cs$_2$CuCl$_4$ and Cs$_{\text 2}$CuBr$_{\text 4}$ from our FTLM
results.
\end{abstract}

\maketitle

\clearpage
\section{Introduction}

The 2D triangular $S=1/2$ Heisenberg antiferromagnet (HAF) is the
archetypical geometrically frustrated quantum magnet~\cite{nakatsuji:10}.
Therefore it has been studied in countless investigations.  Although
there is no pure physical realization due to in-plane symmetry
breaking and inter-plane coupling several compounds fall into the
wider class of anisotropic triangular magnets where the exchange
coupling $J_2$ along one of the triangle sides is different from $J_1$
along the other two sides~\cite{schmidt:14}.  In fact this
generalization is not a drawback but opens interesting possibilities.
It allows to consider a smooth interpolation between square lattice
HAF to isotropic triangular magnet to quasi-1D chain compounds as
function of a single tuning parameter $J_2/J_1$.

Most of the work on this magnetic system was focused on the zero
temperature phase diagram as function of anisotropy ratio, spin wave
excitations and influence of quantum fluctuations as well as effects
of external fields like plateau formation in the magnetization.  We
will not touch these fundamental topics but refer to
Ref.~\cite{schmidt:14} and references cited therein.  Here we will
focus on a more practical issue, namely the determination of the
exchange constant or at least their anisotropy ratio from
thermodynamic quantities of triangular magnets.  This can be done by
analyzing the experimental temperature dependence of specific heat and
magnetic susceptibility where the former is less favorable due to
lattice contributions~\cite{kini:06}.  A simplified but not
unambiguous determination of exchange constants is possible by
identifying position and height of the maximum in thermodynamic
quantities and comparing with theoretical predictions.  Further useful
tools of diagnosing the exchange model are magnetothermal, e.g.
magnetocaloric properties as well as high field magnetization and
saturation field.  This program has sofar mainly been carried out for
frustrated $J_1$-$J_2$ square lattice
magnets~\cite{shannon:04,schmidt:07b,schmidt:10}.

Here we focus on the numerical investigation of much less studied
finite temperature properties of anisotropic triangular quantum spin
systems.  Previous investigations used variants of Monte Carlo methods
for the quantum case~\cite{kawamura:84,suzuki:98,kulagin:13}, also for
the classical case~\cite{melchy:09,tamura:13} or analytical methods
for the low temperature regime~\cite{mezio:12} and high temperature
series expansion~\cite{zheng:05}.  In the present work we use the
finite temperature Lanczos method (FTLM) based on the exact
diagonalization (ED) of finite lattice tiles for this purpose.  The
method is based on evaluating the partition function by averaging over
random starting vectors in the ED procedure.  Using periodic boundary
conditions and discrete symmetries we can go up to largest cluster
size of $N=28$ sites for thermodynamic averages.  We also consider
smaller clusters of size $N=16,20,24$ by FTLM. Furthermore we present
the full analytical solution of the spectrum for the smallest $N=8$
cluster as a reference point.  However, we note that unlike zero
temperature ED results for finite clusters the FTL method cannot be
used for finite size scaling analysis of thermodynamic quantities.
Nonetheless in the special case of the 1D chain system ($J_1=0$) we
can compare to exact results which are in excellent agreement with the
$N=28$ cluster results.  This indicates that our FTLM results are
trustworthy to use as a representation for the thermodynamic limit.

Our main emphasis is the calculation of peak position $T_{\text{max}}$
and height $C_V(T_{\text{max}})$ and $\chi(T_{\text{max}})$ in the
temperature dependence of specific heat and susceptibility,
respectively.  We investigate its systematic variation with anisotropy
control parameter $J_2/J_1$ or $\phi=\tan^{-1}(J_{2}/J_{1})$, in
particular when it is tuned between the simple special cases mentioned
above.  This information is of great importance for the analysis of
thermodynamic data of triangular magnets.  The $T_{\text{max}}(\phi)$
dependence turns out to be considerably more asymmetric than for the
related square lattice model~\cite{shannon:04}.  We show that in
certain ranges of the control parameter an indication of second
maximum or shoulder in these quantities appears.  Furthermore as an
example we discuss the analysis of the full temperature dependence of
$\chi(T)$ for the most prominent triangular quantum spin systems
Cs$_2$CuCl$_4$ and the isostructural Cs$_{\text 2}$CuBr$_{\text 4}$
using FTLM results.  We demonstrate that the derived exchange model is
in excellent agreement with the one obtained from direct spectroscopic
methods like inelastic neutron scattering (INS, Cs$_{\text
2}$CuCl$_{\text 4}$ only) spin wave results and electron spin
resonance (ESR) experiments, both in high fields.

In Sec.~\ref{sec:model} we define the model and its parametrization.
The reference of the full solution for the eight-site cluster is
discussed in Sec.~\ref{eight-site}.  The FTL method and its technical
implementation are discussed to some extent in Sec.~\ref{finitetemp}.
Our main results on specific heat and susceptibility are presented in
Secs.~\ref{sec:heatcapacity} and \ref{sec:susceptibility},
respectively and the explicit comparison to Cs$_2$CuCl$_4$ and
Cs$_2$CuBr$_4$ is discussed in Sec.~\ref{CsCu}.  Finally
Sec.~\ref{conclusion} gives the conclusion and outlook.

\section{Anisotropic triangular exchange model and its parametrization}
\label{sec:model}

\begin{figure}
    \includegraphics[width=0.78\columnwidth]{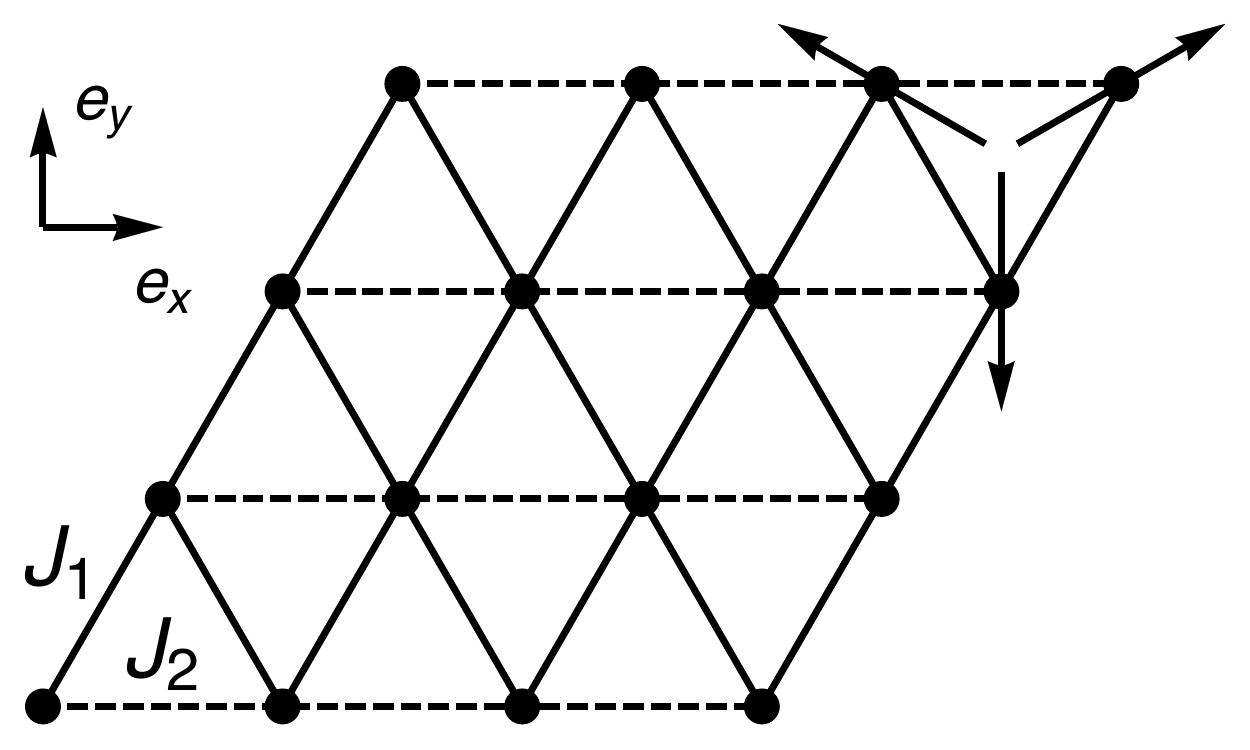}\hfill
    \caption{Schematic view of anisotropic triangular exchange models.
    For $J_1=0$ ($\phi=\pm\pi/2$) this reduces to 1D $J_2$-spin chains
    ($\parallel$), for $J_2=0$ ($\phi=0$) to the square lattice $J_1$-
    HAF ($\square$) and for $J_1=J_2$ ($\phi=\pi/4$) to the isotropic
    triangular system ($\triangle$).}
    \label{fig:structure}
\end{figure}
\begin{figure}
    \includegraphics[width=\columnwidth]{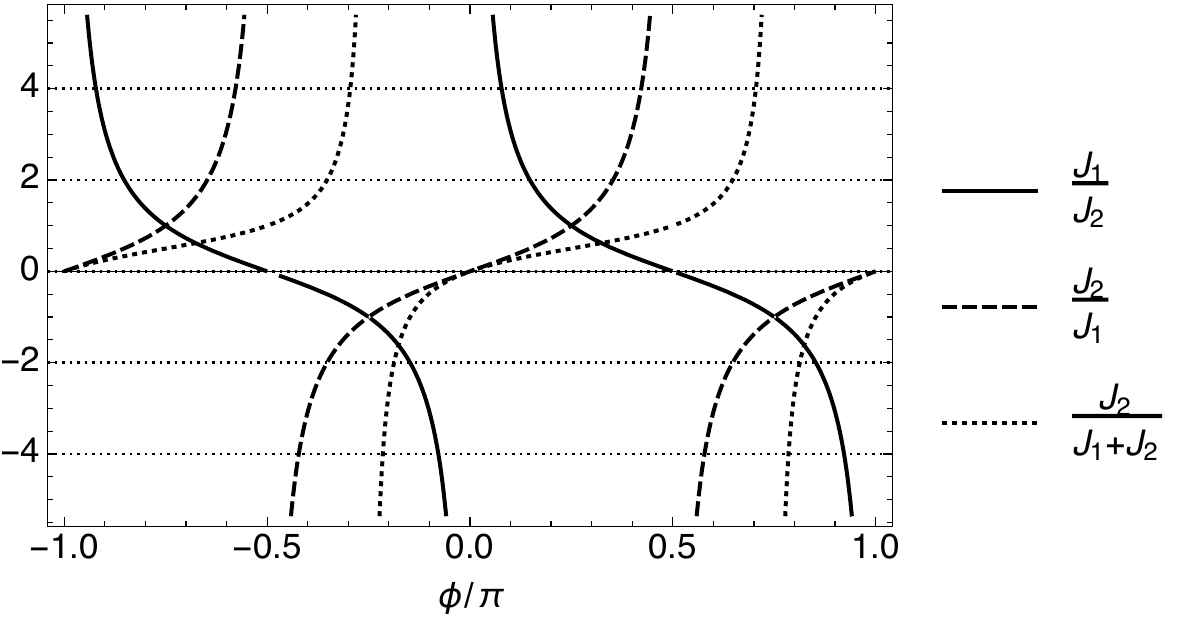}
    \caption{Different exchange parametrizations used for the
    anisotropic triangular Heisenberg model and their dependence on the
    anisotropy parameter $\phi$.  The left and right branch for
    $J_2/J_1$ corresponds to $J_2<0$ or $J_2 >0$, respectively.}
    \label{fig:par}
\end{figure}

The  anisotropic  $J_{1}-J_{2}$ exchange model on the triangular lattice 
(Fig.~\ref{fig:structure}) is given by
\begin{equation}
    {\cal H}
    =
    \sum_{\left\langle ij\right\rangle}J_{ij}\vec S_{i}\cdot\vec S_{j}
    \label{eqn:h}
\end{equation}
with
\begin{equation}
    J_{ij}
    =
    \left\{
    \begin{array}{l@{\mbox{ if }}l}
        J_{1} & \vec R_{j}=
        \vec R_{i}\pm\frac 12\left(
        \vec e_{x}\pm\sqrt{3}\vec e_{y}\right)\\
        J_{2} & \vec R_{j}=
        \vec R_{i}\pm\vec e_{x}
    \end{array}
    \right.,
\end{equation}
where $\vec e_{x}, \vec e_{y}$ are unit vectors along cartesian
directions.  Figure \ref{fig:structure} may also be interpreted as
tilted square lattice model with one diagonal $J_2$ bond cut
out~\cite{schmidt:14}.  For the anisotropic triangular-lattice model,
different parametrizations of the exchange energies are customary.  As
introduced in Ref.~\cite{shannon:04} and used subsequently we prefer
the polar representation defined by
\begin{eqnarray}
    J_{1}=J_{\text c}\cos\phi,
    &\quad&
    J_{2}=J_{\text c}\sin\phi,
    \\
    J_{\text c}=\sqrt{J_{1}^{2}+J_{2}^{2}},
    &\quad&
    \phi=\tan^{-1}\left(\frac{J_{2}}{J_{1}}\right)
    \nonumber
\end{eqnarray}
where $J_{\text c}$ gives the overall energy scale and $\phi$ is the
anisotropy control parameter.  This parametrization allows for an easy
interpolation between important geometrical limiting cases, namely the
square-lattice N\'eel antiferromagnet with $J_{2}=0$ ($\phi=0$), the
isotropic $120^\circ$ triangular antiferromagnet with $J_{2}=J_{1}$
($\phi=\pi/4$), the antiferromagnetic chain with $J_{1}=0$
($\phi=\pi/2$), and their ferromagnetic counterparts.  However, there
are alternative possibilities in the literature which we briefly
mention here for ease of comparison (Fig.~\ref{fig:par}).  Many
authors regard the model as an extension to the one-dimensional spin
chain, therefore they use the exchange $J_{2}$ along these chains as
the overall energy unit and parametrize their results in terms of
$\alpha:=J_{1}/J_{2}$ with $J_{1}$ being the interchain coupling.
From a square-lattice perspective in turn, the exchange parameter
$J_{1}$ appears to be the natural energy unit, and correspondingly
$1/\alpha:=J_{2}/J_{1}$ is used as well.

Both model parameters $\alpha$ and $1/\alpha$ however are problematic
when trying to describe the full phase diagram, in particular the
interpolation between the square-lattice antiferromagnet
($\alpha\to\infty$) and the one-dimensional chain
($1/\alpha\to\infty$).  To overcome this, the function
$f:=J_{2}/(J_{1}+J_{2})=1/(1+\alpha)$ has been introduced in
Ref.~\cite{zheng:99}, which remains finite in both limits and
between these.

However the latter parametrization also does not cover the full phase
diagram unambiguously, which is why we have introduced the universal
energy scale $J_{\text c}$ and the anisotropy angle $\phi$.  In the
square-lattice case, this has the additional advantage that $J_{\text
c}$, apart from a global factor $\sqrt2$, denotes the magnetocaloric
energy scale $J_{\text{mc}}$ as well~\cite{shannon:04}.  To facilitate
the comparison of our results with the literature, Fig.~\ref{fig:par}
displays the dependence of the quantities introduced here on the
anisotropy control parameter $\phi$.

\begin{figure}
    \includegraphics[width=.8\columnwidth]{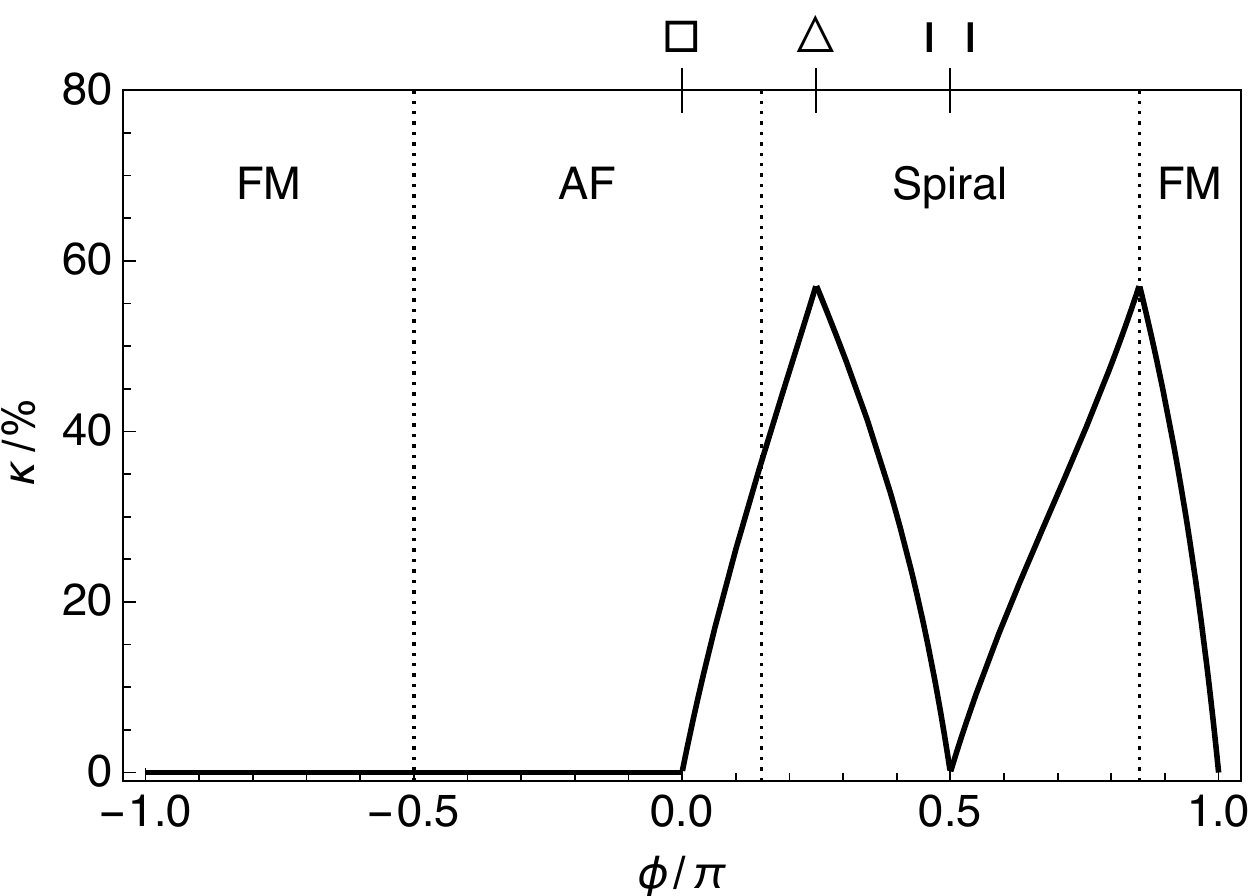}
    \caption{Degree of frustration in per cent in the triangular HAF
    as function of exchange anisotropy parameter.  It vanishes
    identical in the unfrustrated regime $\phi<0 \;(J_2<0)$.  The
    dotted lines are the classical phase boundaries~\cite{schmidt:14}.
    }
    \label{fig:frustratio}
\end{figure}
The nearest neighbor (n.n.) HAF model on the triangular lattice is the
generic `geometrically frustrated' spin system where the exchange
energy of n.n.~bonds cannot be minimized simultaneously for all bonds
(see upper corner of Fig.~\ref{fig:structure}).  It is worthwhile to
quantify this intuitive notion of `frustration'.  A measure for it is
the total loss of exchange energy due to frustration relative to the
exchange energy without it.  For the basic triangular (three-site)
plaquette in Fig.~\ref{fig:structure} the degree of frustration is
then given by
\begin{equation}
    \kappa(\phi):=1-\frac{E_\triangle}{E_{\text t}+E_{\text d}}
    \label{eq:kappa}
\end{equation}
Here $E_\triangle$ is the ground state energy of the frustrated
triangle. $E_{\text t}:=E_{\triangle}(J_{2}=0)$ and $E_{\text
d}:=E_{\triangle}(J_{1}=0)$ are the ground state energies of its
unfrustrated trimer and dimer parts, respectively with
\begin{equation}
    E_\triangle(\phi)
    :=
    \min\left(
    -\frac{3}{4}J_2,-J_1+\frac{1}{4}J_2,\frac{1}{2}J_1+\frac{1}{4}J_2
    \right),
\end{equation}
where the minimum is taken from the three different energy eigenvalues
of the triangle.  Using these expressions in Eq.~(\ref{eq:kappa}) the
degree of frustration (in per cent) in the triangular HAF as function
of control parameter is shown in Fig.~\ref{fig:frustratio}.  There the
dotted lines indicate the classical boundaries between FM, AF and
spiral phases as discussed in Ref.~\cite{schmidt:14}.  $\kappa(\phi)$
vanishes identically in the unfrustrated ($J_2<0$ or $\phi<0$) case.
In the frustrated regime ($J_2>0$ or $\phi>0$) it achieves its maximum
values of $\kappa =4/7$ corresponding to $57\,\%$ frustration at the
isotropic point ($\triangle$) and the Spiral/FM phase boundary while
it vanishes for the 1D chain case ($\parallel$) where $J_1=0$.  We
note that the reduction in the ordered ground state moment in the
frustrated regime does not directly follow the degree of frustration
but is the result of the subtle interplay of the latter with quantum
fluctuations~\cite{schmidt:14}.

\section{Precursor: classical phase diagram and eight-site full solution}
\label{eight-site}

Within the finite temperature Lanczos method the thermodynamic
quantities will be computed directly using an averaging procedure over
the low energy spectrum only.  To obtain a better intuition for the
triangular Heisenberg model it is useful to calculate also the full
spectrum by direct diagonalization of small clusters.  We will
demonstrate that it develops characteristic signatures at the
classical phase boundaries and special symmetry points as function of
control parameter using the eight-site cluster.

\subsection{Spectrum and classical phase diagram}

\begin{figure}
    \centering
    \includegraphics[width=.5\columnwidth]{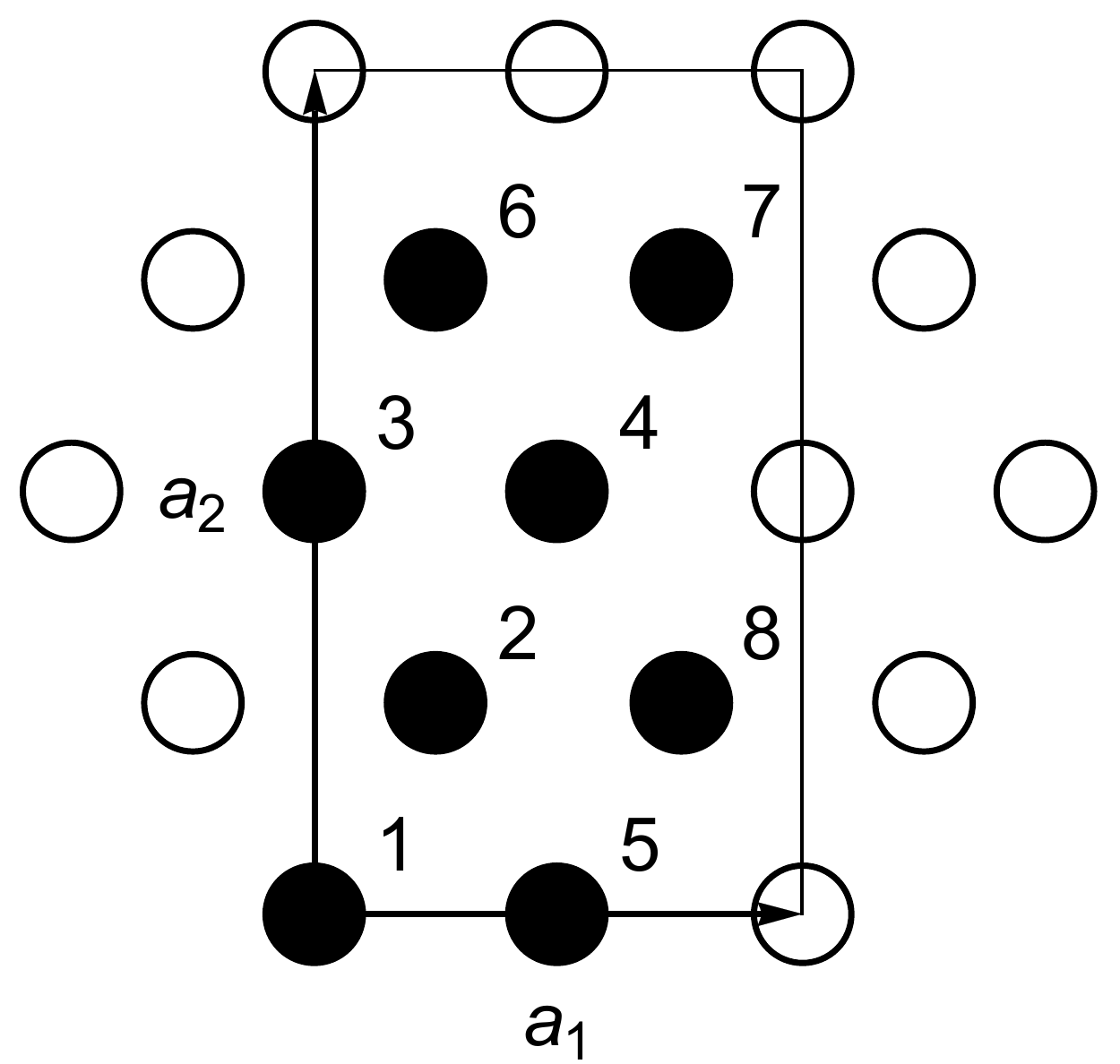}
    \caption{Sketch of the tile 8:2-2. The edge vectors are $\vec 
    a_{1}$ and $\vec a_{2}$, the numbers label the individual sites.}
    \label{fig:eight-site-tile}
\end{figure}
As discussed in Ref.~\cite{shannon:04} for the square-lattice model,
we can express the spectrum of the triangular eight-site cluster on
tile 8:2-2 with periodic boundary conditions as a sum over
Hamiltonians on complete graphs.  We define
\begin{equation}
    {\cal H}_{\cal L}^{\text{CG}}
    :=
    \sum_{i<j\in{\cal L}}
    \vec S_{i}\cdot\vec S_{j}
    =
    \frac12\left[
    \left(\sum_{i\in\cal L}\vec S_{i}\right)^{2}
    -\sum_{i\in\cal L}\vec S_{i}^{2}
    \right]
\end{equation}
where ${\cal L}$ denotes an ordered list of lattice sites. In our
case it has either two elements (a bond), four elements (a square or
tetrahedron), or eight elements (a cube).  We then can write
\begin{eqnarray}
    {\cal H}_{8}
    &=&
    J_{1}\left[
    {\cal H}_{\{1,2,3,4,5,6,7,8\}}^{\text{CG}}
    -\left(
    {\cal H}_{\{1,3,4,5\}}^{\text{CG}}+{\cal H}_{\{2,6,7,8\}}^{\text{CG}}
    \right)
    \right]
    \nonumber\\&&{}
    +2J_{2}\left[
    {\cal H}_{\{1,5\}}^{\text{CG}}
    +{\cal H}_{\{3,4\}}^{\text{CG}}
    +{\cal H}_{\{2,8\}}^{\text{CG}}
    +{\cal H}_{\{6,7\}}^{\text{CG}}
    \right]
    \label{eqn:height}
\end{eqnarray}
for the Hamiltonian.  The tile and its labelling is illustrated in
Fig.~\ref{fig:eight-site-tile}.  From Eq.~(\ref{eqn:height}) with
$S=1/2$ being the maximum expectation value for each local spin $\vec
S_{i}$, we get the corresponding list of eigenvalues by hierarchically
constructing all possible multiple-spin configurations starting with
the basic two-spin singlets and triplets on the pairs $\{1,5\}$,
$\{3,4\}$, $\{2,8\}$, and $\{6,7\}$: From all pairs of two-spin
states, we can construct the four-spin states with total spin
$S=0,1,2$, each possible pair of these four-spin states in turn can be
combined to an eight-spin state with total spin $S=0,1,2,3,4$ and spin
degeneracy $(2S+1)$ .  Because the total spin may be composed in
several ways by the spins of sub-clusters there exist additional
degeneracies.  In this way, we can construct a total of ${N\choose
N/2}=70$ states for the eight-site cluster.
    
\begin{table}
    \caption{All energy levels of the eight-site cluster.  $S_{1-8}$
    is the total cluster spin and the next six columns give the
    sub-cluster spins.  The last column denotes the additional
    degeneracy of levels on top of the $(2S_{1-8}+1)$-fold spin
    degeneracy.}
    $$
    \begin{array}{c|cccccc|c|c}
        S_{1-8} & S_{1345} & S_{15} & S_{34} & S_{2678} & S_{28} & S_{67} &
        \text{energy} &
        \text{deg.} \\
        \hline
        0 & 0 & 0 & 0 & 0 & 0 & 0 & -6 J_2 & \;\;1* \\
        0 & 0 & 0 & 0 & 0 & 1 & 1 & -2 J_2 & 2 \\
        0 & 0 & 1 & 1 & 0 & 1 & 1 & 2 J_2 & 1 \\
        0 & 1 & 0 & 1 & 1 & 0 & 1 & -2 \left(J_1+J_2\right) & 4 \\
        0 & 1 & 0 & 1 & 1 & 1 & 1 & -2 J_1 & 4 \\
        0 & 1 & 1 & 1 & 1 & 1 & 1 & 2 J_2-2 J_1 & 1 \\
        0 & 2 & 1 & 1 & 2 & 1 & 1 & 2 J_2-6 J_1 &\; \;1* \\
        1 & 0 & 0 & 0 & 1 & 0 & 1 & -4 J_2 & 4 \\
        1 & 0 & 0 & 0 & 1 & 1 & 1 & -2 J_2 & 2 \\
        1 & 0 & 1 & 1 & 1 & 0 & 1 & 0 & 4 \\
        1 & 0 & 1 & 1 & 1 & 1 & 1 & 2 J_2 & 2 \\
        1 & 1 & 0 & 1 & 1 & 0 & 1 & -J_1-2 J_2 & 4 \\
        1 & 1 & 0 & 1 & 1 & 1 & 1 & -J_1 & 4 \\
        1 & 1 & 1 & 1 & 1 & 1 & 1 & 2 J_2-J_1 & 1 \\
        1 & 1 & 0 & 1 & 2 & 1 & 1 & -3 J_1 & 4 \\
        1 & 1 & 1 & 1 & 2 & 1 & 1 & 2 J_2-3 J_1 & 2 \\
        1 & 2 & 1 & 1 & 2 & 1 & 1 & 2 J_2-5 J_1 & 1 \\
        2 & 0 & 0 & 0 & 2 & 1 & 1 & -2 J_2 & 2 \\
        2 & 0 & 1 & 1 & 2 & 1 & 1 & 2 J_2 & 2 \\
        2 & 1 & 0 & 1 & 1 & 0 & 1 & J_1-2 J_2 & 4 \\
        2 & 1 & 0 & 1 & 1 & 1 & 1 & J_1 & 4 \\
        2 & 1 & 1 & 1 & 1 & 1 & 1 & J_1+2 J_2 & 1 \\
        2 & 1 & 0 & 1 & 2 & 1 & 1 & -J_1 & 4 \\
        2 & 1 & 1 & 1 & 2 & 1 & 1 & 2 J_2-J_1 & 2 \\
        2 & 2 & 1 & 1 & 2 & 1 & 1 & 2 J_2-3 J_1 & 1 \\
        3 & 1 & 0 & 1 & 2 & 1 & 1 & 2 J_1 & 4 \\
        3 & 1 & 1 & 1 & 2 & 1 & 1 & 2 \left(J_1+J_2\right) & 2 \\
        3 & 2 & 1 & 1 & 2 & 1 & 1 & 2 J_2 & 1 \\
        4 & 2 & 1 & 1 & 2 & 1 & 1 & 2 \left(2 J_1+J_2\right) & \;\; 1 *\\
    \end{array}
    $$
    \label{tbl:spectrum}
\end{table}
\begin{figure}
    \centering
    \includegraphics[width=\columnwidth]{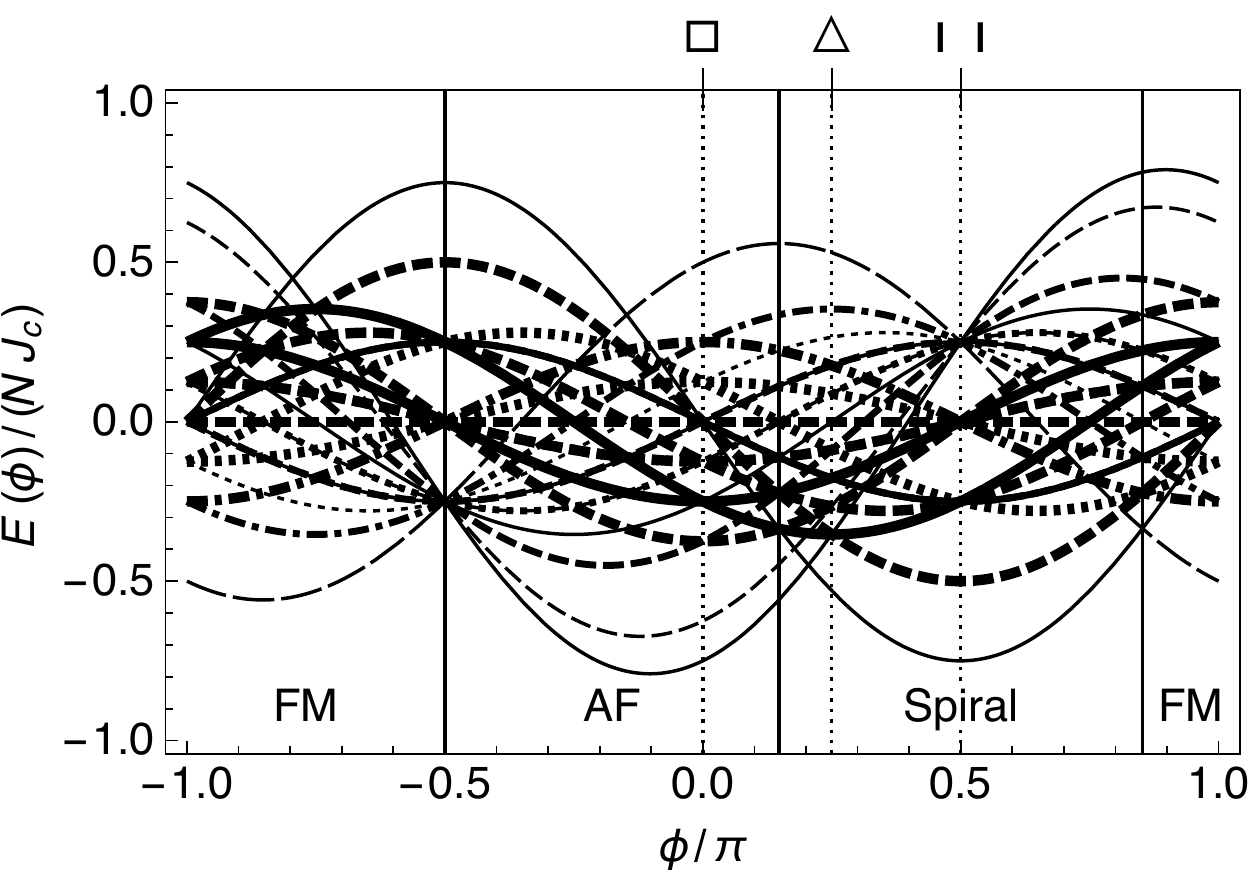}
    \caption{Energy spectrum of the eight-site cluster, classified
    according to total spin $S$.  Solid lines: $S=0$, short-dashed
    lines: $S=1$, dotted lines: $S=2$, dash-dotted lines: $S=3$,
    long-dashed lines: $S=4$.  The line thickness indicates the
    degeneracy of the corresponding energy level, see
    Table~\ref{tbl:spectrum}.  The thin vertical lines denote the
    classical phase boundaries between ferromagnet (FM), Néel
    antiferromagnet (AF), and spiral structure.  The dotted vertical
    lines denote the three special cases square-lattice Néel
    antiferromagnet ($J_{1}>0$, $J_{2}=0$), isotropic triangular lattice
    ($J_{2}\equiv J_{1}>0$), and one-dimensional antiferromagnetic
    chains ($J_{1}=0$, $J_{2}>0$).}
    \label{fig:spectrum}
\end{figure}
Table~\ref{tbl:spectrum} displays the complete set of eigenvalues,
together with their (additional) degeneracies, total spins, and total
spins on the sub-lattices.  Of particular interest are those states
corresponding to classical ground states (marked by an asterisk in the
last column): the columnar antiferromagnet (first state in the table,
energy $-6J_{2}$), the Néel antiferromagnet (seventh state, energy
$2J_{2}-6J_{1}$), and the ferromagnet (last state, energy
$2(2J_{1}+J_{2})$.  These three states replace each other as ground
states as a function of the anisotropy ratio $\phi$, see
Fig.~\ref{fig:spectrum}.  Fig.~\ref{fig:spectrum} displays the full
spectrum of the tile 8:2-2 as a function of the anisotropy angle.  The
thicknesses of the lines indicate the degeneracies of the
corresponding states as listed in Table~\ref{tbl:spectrum}.  Solid
lines denote singlet states, the long-dashed line denotes the
ferromagnet, for further line types see the figure caption.

We overlay this spectrum to the classical phase diagram of the model
discussed in detail in Ref.~\cite{schmidt:14}.  The classical FM, AF
and spiral phases are separated by the thin vertical lines.  The thin
dotted lines indicate as special cases the square-lattice Néel
antiferromagnet ($\square$, $\phi=0$ or $J_{1}>0$, $J_{2}=0$),
isotropic antiferromagnetic triangular lattice ($\triangle$,
$\phi=\pi/4$ or $J_{2}\equiv J_{1}>0$) corresponding to the
non-collinear $120^\circ$ commensurate spiral structure and
one-dimensional antiferromagnetic chains ($\parallel$, $\phi=\pi/2$ or
$J_{1}=0$, $J_{2}>0$).

Characteristic for the spectrum at  these six special points is the high number of
degenerate excited states with different total spin.  At the borders
of the classical ferromagnetic phase also the ground state changes
from the fully polarized state to a singlet.  However the classical
phase boundary between the antiferromagnet and the spiral phase does
not correspond to an equivalent change of (in this case nonmagnetic)
ground states.  This is due to the fact that with our eight-site tile
we cannot approximate any incommensurate state at all, therefore the
``best'' approximation to the ``true'' ground state remains the Néel
state for $1/2\le J_{2}/J_{1}\le3/4$ (equivalent to
$0.148\le\phi/\pi\le0.205$), and the columnar antiferromagnetic state
for larger values of $J_{2}/J_{1}$.

Similar to the square-lattice case~\cite{shannon:04}, we observe level
crossings near the classical phase boundary between AF and spiral
phase.  For the square lattice, we took this as an indication for the
appearance of a nonmagnetic phase, coinciding with the suppression of
the ordered moment in a region around the classical transition
observed within linear spin-wave theory~\cite{schmidt:14}.
Qualitatively the same happens here, however this analogy should not
be taken too far, because no indication of that kind exists for the
classically invisible crossover to one-dimensional chains at
$\phi=\pi/2$, which is in reality surrounded by a large nonmagnetic
region.  And empirically we know that linear spin-wave theory rather
underestimates the size of these nonmagnetic regions.  At the
crossover from the spiral to the ferromagnetic phase for
$\phi/\pi\approx0.852$, a similar type of pattern of level crossings
exists.  Linear spin-wave theory gives inconsistent results in this
case.

\subsection{Thermodynamics of the eight-site cluster}

Given the eigenvalue $E_{I}$ with degeneracy $d_{I}$ and total spin
$S_{I}$ for each state $I$ listed in Table~\ref{tbl:spectrum}, we can
easily evaluate the partition function ${\cal Z}=
\sum_{I}d_{I}\left(2S_{I}+1\right){\rm e}^{-\beta E_{I}}$ to calculate
the magnetic susceptibility $\chi$ and the specific heat $c_{V}$
according to
\begin{eqnarray}
    \chi
    &=&
    \frac{\beta J_{\text c}}N
    \frac{1}{\cal Z}\sum_{I}
    d_{I}(2S_{I}+1)\frac{S_{I}(S_{I}+1)}3
    {\rm e}^{-\beta E_{I}},
    \label{eqn:chieight}
    \\
    c_{V}
    &=&
    \frac{\beta^{2}}{N}\left[
    \frac{1}{\cal Z}\sum_{I}
    d_{I}(2S_{I}+1)E_{I}^{2}{\rm e}^{-\beta E_{I}}
    \right.
    \nonumber\\
    &\phantom{=}&
    \hphantom{\frac{\beta^{2}}{N}}
    \left.{}
    -\left(
    \frac{1}{\cal Z}\sum_{I}
    d_{I}(2S_{I}+1)E_{I}{\rm e}^{-\beta E_{I}}
    \right)^{2}
    \right]
    \label{eqn:cveight}
\end{eqnarray}
where $\beta=1/(k_{\text B}T)$ with $k_{\text B}$ being the Boltzmann
constant.  Here and in the following, we express the molar
susceptibility in units of $N_{\text L}\mu_{0}\left(g\mu_{\text
B}\right)^{2}J_{\text c}^{-1}$, where $N_{\text L}$ is the Loschmid
constant, $\mu_{0}$ the magnetic permeability constant, $g$ the
gyromagnetic ratio, and $\mu_{\text B}$ the Bohr magneton.  For the
dimension of the specific heat $c_{V}$, we use the universal gas
constant $R=N_{\text L}k_{\text B}$.

\section{General remarks on finite temperature methods}
\label{finitetemp}

The finite temperature properties of spin systems can be treated with
analytical high temperature
expansions~\cite{zheng:05,oitmaa:06,rosner:03,nath:08,schmidthj:11} or
with the numerical FTL method~\cite{jaklic:00} which will be employed
in this work.  As a reference we first discuss briefly the single
first order term of the expansion method.

\subsection{High temperature approximation}

The high-temperature behavior of the susceptibility $\chi$ and the
specific heat $c_{V}$ to quadratic order in $\beta$ is determined by 
the Curie-Weiss energy $\Theta$ and the magnetocaloric energy scale 
$J_{\text{mc}}$, which are defined through
\begin{eqnarray}
    \Theta&:=&\frac{S(S+1)}3\sum_{n}J_{ii+n}
    =J_{1}+\frac{J_{2}}2,
    \\
    J_{\text{mc}}^{2}&:=&\frac12\sum_{n}J_{ii+n}^{2}
    =2J_{1}^{2}+J_{2}^{2}.
    \label{eqn:jth}
\end{eqnarray}
The sums run over all bonds $n$ connecting an arbitrary but fixed site
$i$ with its (not necessarily nearest) neighbors at sites $\{i+n\}$.
We then have~\cite{schmidthj:11}
\begin{eqnarray}
    \chi
    &=&
    \frac{S(S+1)}3\beta J_{\text c}\left(1-\beta\Theta\right),
    \\
    c_{V}
    &=&
    \frac13\left[S(S+1)\right]^{2}\beta^{2}J_{\text{mc}}^{2}.
    \label{eqn:cvht}
\end{eqnarray}
In principle we should be able to determine the exchange parameters
$J_{1}$ and $J_{2}$ already from high-temperature fits of the experimental
results to the expressions above.  However having fixed
$J_{\text{mc}}^{2}$ and $\Theta$ determines $2J_{1}+J_{2}$ and
$|J_{1}-4J_{2}|$ but leaves the sign of the latter undetermined.  When
expressed with the parameters $J_{\text c}$ and $\phi$, this is
equivalent to the fact that there are, apart from special cases,
always {\em two\/} possible values $\phi_{\pm}$ for the anisotropy
parameter.  These two values $\phi_{\pm}$, however, can lie in two
different thermodynamic phases with completely different properties. This
ambiguity is similar to the one in the square lattice  $J_{1}-J_{2}$ model 
and its implications there were discussed in 
Refs.~\onlinecite{misguich:03,shannon:04}.

Although the coefficients of the high-temperature expansions for
$\chi(T)$ and $c_{V}(T)$, being polynomial functions of $J_{1}$ and
$J_{2}$, are known up to at least eighth order~\cite{schmidthj:11},
this situation essentially will not change by including further
higher-order terms in a high-temperature expansion~\cite{misguich:03},
and it remains difficult to determine $J_{1}$ and $J_{2}$
unambiguously solely from fits to the high temperature dependence of $\chi$ and
$c_{V}$.  One powerful further diagnostic is the investigation
of saturation fields~\cite{schmidt:08,schmidt:10} provided that they
are in an accessible range.

\begin{figure*}
    \centering
    \includegraphics[width=.2\textwidth]{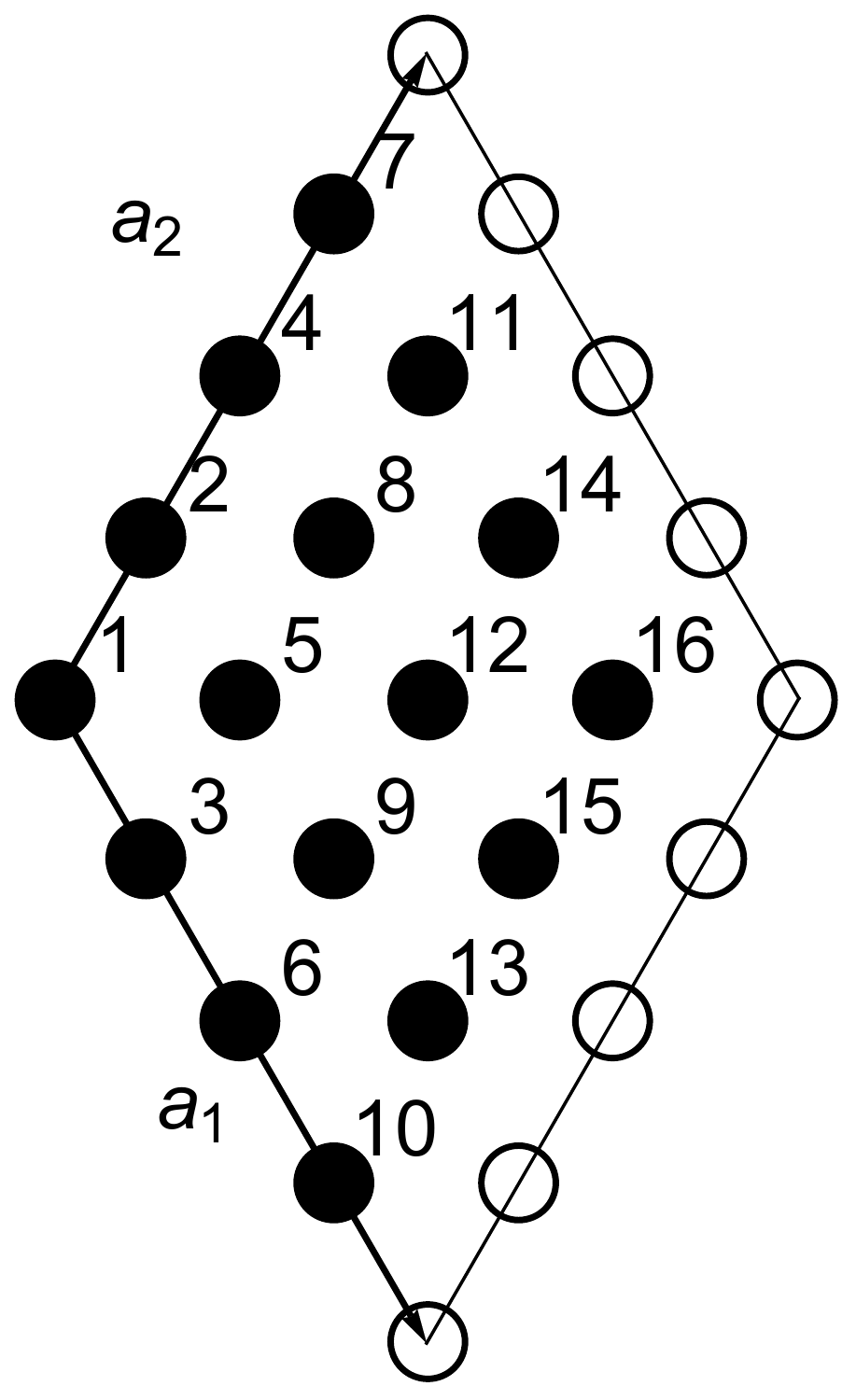}
    \includegraphics[width=.25\textwidth]{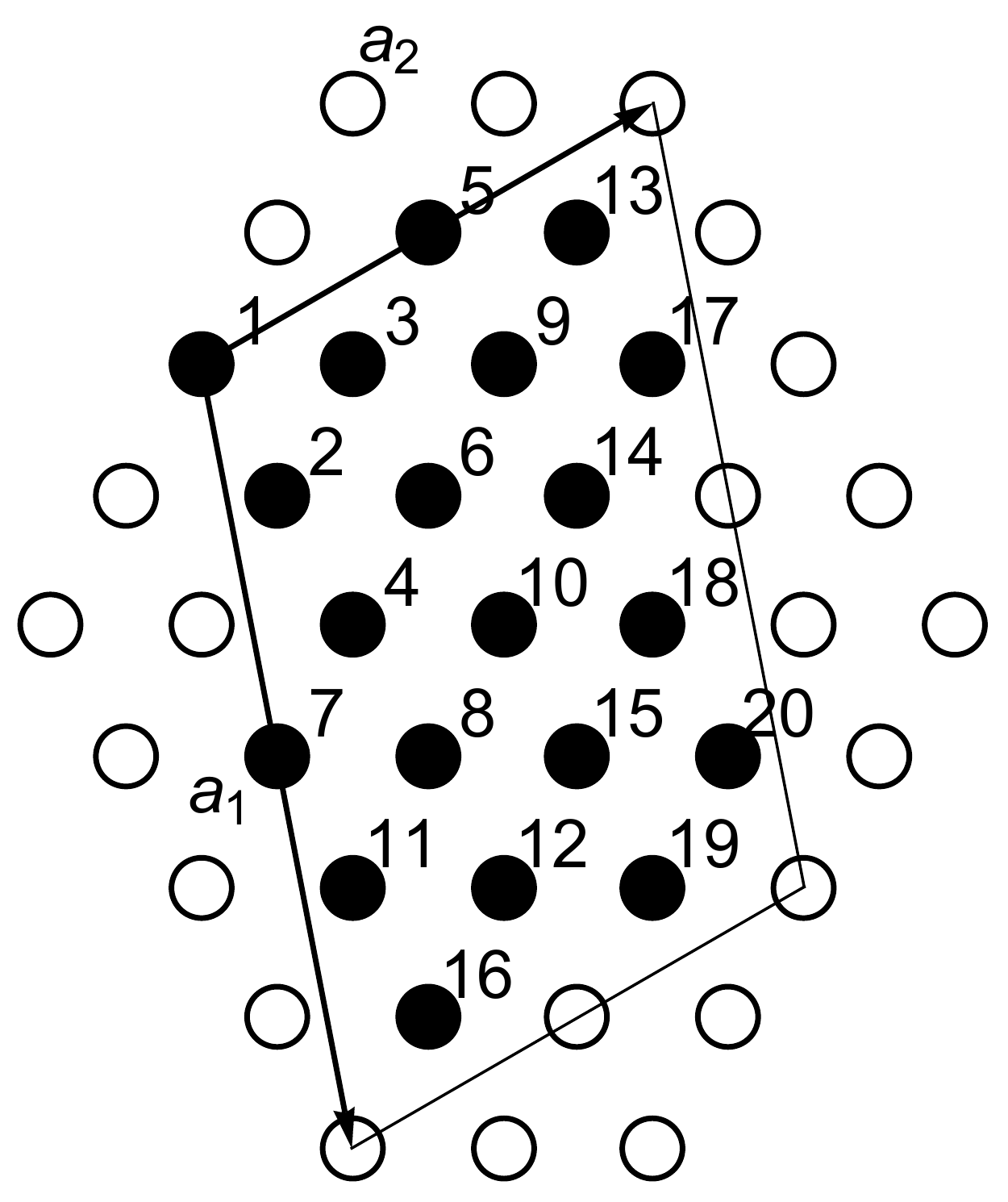}
    \includegraphics[width=.2\textwidth]{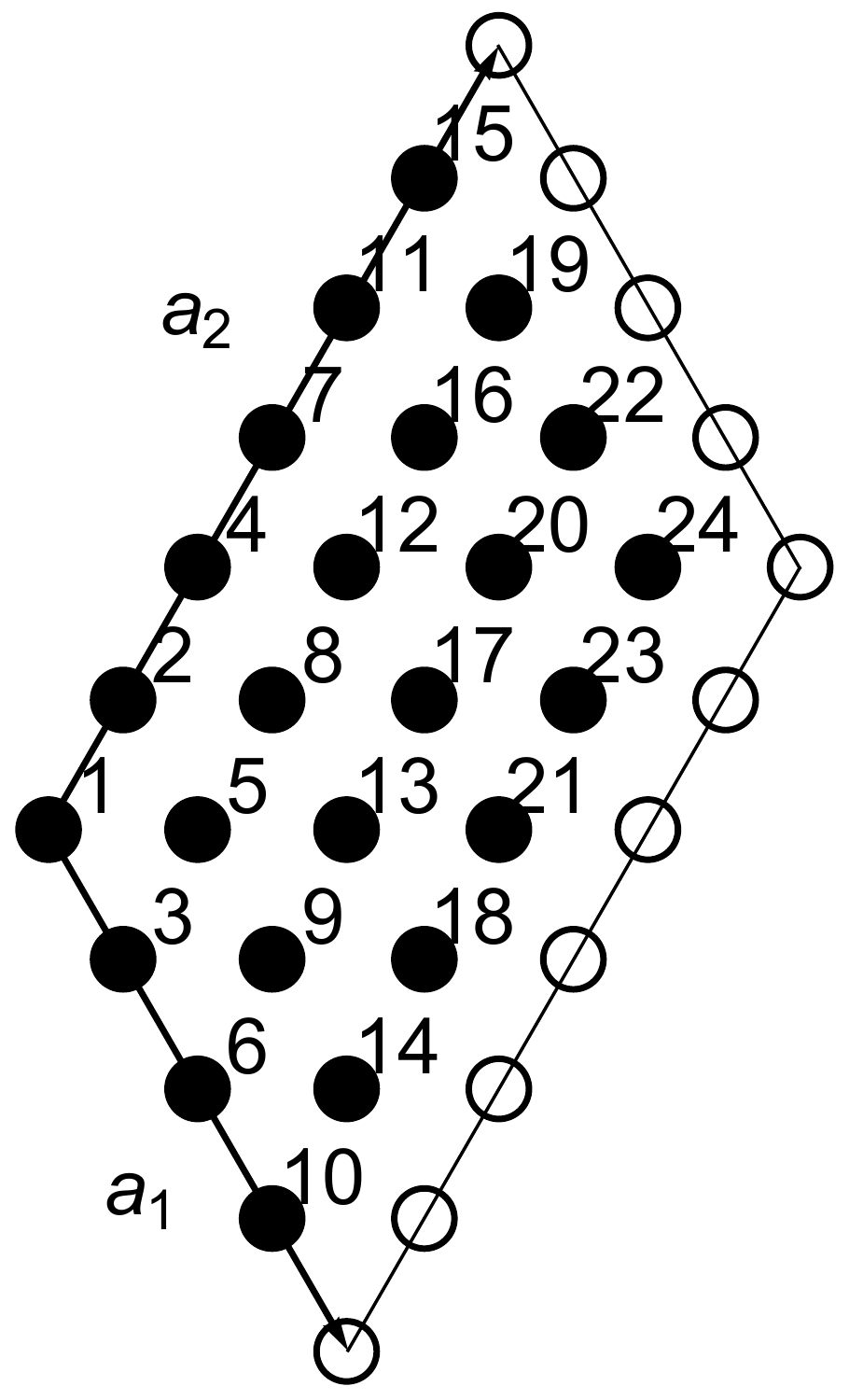}
    \includegraphics[width=.25\textwidth]{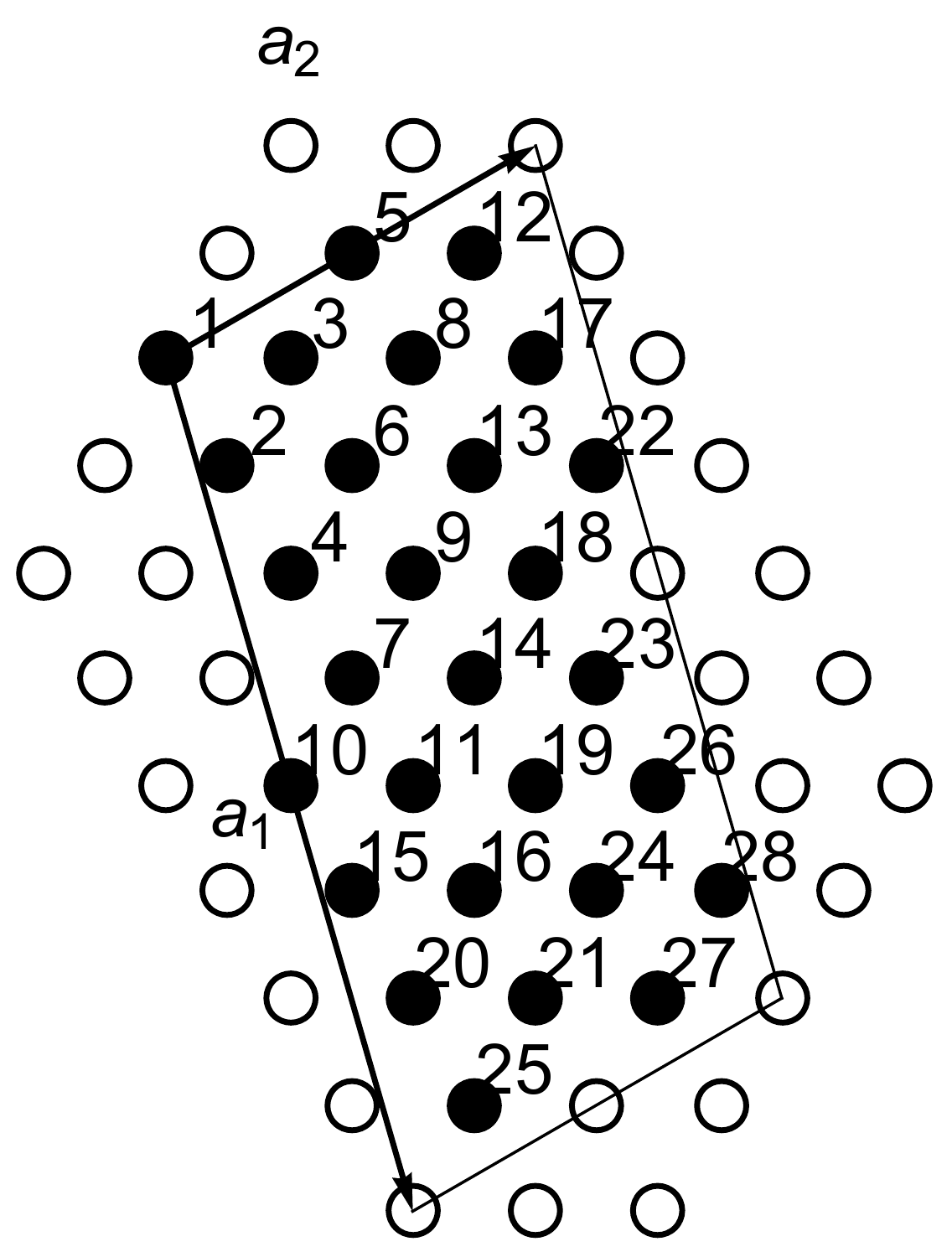}
    \caption{Tiles used in the finite-temperature Lanczos 
    calculations. From left to right: 16:4-0, 20:2-4, 24:4-0, 28:2-4. 
    The numbers label the lattice sites, $\vec a_{1}$ and $\vec 
    a_{2}$ are the edge vectors.}
    \label{fig:tiles}
\end{figure*}
\subsection{Finite-temperature Lanczos method}

To overcome this ambiguity, we use the finite-temperature Lanczos
method~\cite{hanebaum:14,schmidt:07b,jaklic:00} to evaluate the
thermodynamic functions directly and compare specific heat and
susceptibility temperature dependence over the whole temperature range
above the finite size gap region.  The method is based on the
evaluation of thermodynamic traces using the eigenvalues and
many-particle wave functions determined by numerical exact
diagonalization of the Hamiltonian matrix on finite tiles.  After
mapping the Hamiltonian onto a sparse matrix, we use the iterative
Lanczos algorithm~\cite{lanczos:50} to generate the first few (between
$1$ and $100$) extremal eigenvalues and the corresponding wave
functions.  Due to the Boltzmann weight, these are the most important
eigenvalues contributing to the partition function.

We classify our wave functions according to the expectation value of
the $z$ component $\Omega_{z}=\sum_{i=1}^{N}S_{i}^{z}$ of the total
spin $\vec\Omega$, the crystal momenta $\vec k$ and the point group
symmetries of the tile.  This brings the Hamiltonian matrix into block
diagonal form, allowing us to go to tile sizes up to $N=28$ in our
finite-temperature calculations.  In this way we can evaluate
thermodynamic traces on industry standard computer hardware.

The tiles we are using for the diagonalization procedures are
illustrated in Fig.~\ref{fig:tiles}.  We apply the same labelling as
described in Ref.~\cite{schmidt:11} for the square-lattice case,
adapted to the triangular lattice.  There are two important
differences: At finite temperatures we cannot easily perform any kind
of finite-size scaling analysis of the partition function, therefore we directly
take the results of the different lattice tilings. In addition
there exists at least one phase corresponding to a classical phase
with an incommensurate ordering vector.  Due to the finiteness of our
$\vec k$ space grid, we cannot model this closely. However for 
thermodynamic properties like susceptibility and heat capacity all 
thermally populated states contribute, and the exact modeling of the 
ground state as a function of our parameters is of secondary 
importance. For zero temperature ED approach the introduction of
twisted boundary conditions may provide a way to circumvent this
problem~\cite{thesberg:14}.

We attempt to determine the thermal expectation value of an arbitrary
static operator $A$ by the fundamental traces over the statistical 
operator,
\begin{eqnarray}
    \left\langle A\right\rangle_{\beta}
    &=&
    \frac{1}{\cal Z}
    \sum_{n=1}^{N_{\text{st}}}\left\langle n\left|
    {\rm e}^{-\beta {\cal H}}A\right|n\right\rangle
    \label{eqn:lancz:trace},
    \\
    {\cal Z} &=& \sum_{n=1}^{N_{\text{st}}}\left\langle n\left| {\rm
    e}^{-\beta {\cal H}}\right|n\right\rangle,
\end{eqnarray}
where $\cal Z$ is the partition function and $N_{\text{st}}$ is the
dimension of the Hilbert space spanned by the basis
$\left\{\left|n\right\rangle:n=1\ldots N_{\text{st}}\right\}$
($N_{\text{st}}\approx2.7\cdot 10^8 $ for $N=28$).  In each
symmetry-invariant subspace of the full Hilbert space, the Lanczos
algorithm in principle is a sophisticated iterative basis change
transforming the original Hamiltonian $\cal H$ to an equivalent one
for an open one-dimensional chain (not ring) problem with complex
on-site potentials and nearest-neighbor interactions.  Each iteration
step corresponds to adding an additional site to this chain.  After
$M$ steps, the resulting equivalent tridiagonal Hamiltonian matrix
${\cal H}_{M}$ can be diagonalized easily to get the eigenvalues
$\left\{\epsilon_{j}:j=1\ldots M\right\}$ and normalized wave
functions $\left\{\left|\psi_{j}\right\rangle:j=1\ldots M\right\}$
such that we can, in principle, evaluate Eq.~(\ref{eqn:lancz:trace}).
The moduli of the additional matrix elements of ${\cal H}_{M}$ rapidly
decrease with increasing number of iterations $M$, a property which we
use as a convergence criterion.  Furthermore the difference in
ground-state energy for iteration $M$ and $M-1$ is used as a second
convergence criterion.  Typical values such that machine precision is
reached for the ground state energy are $10\le M\le100$, which is
vanishingly small compared to the original dimension
$N_{\text{st}}={\cal O}\left(10^{8}\right)$ of the Hilbert space.  To
sample an as large as possible part of the Hilbert space, we start
$N_{\text R}={\cal O}(100)$ iterations with different random wave
functions or starting vectors $|r\rangle$, such that eventually we use
no more than ${\cal O}\left(10^{4}\right)$ eigenvalues and wave
functions per Hamiltonian block.  It may be shown~\cite{jaklic:00}
that this procedure yields an asymptotically exact result.  In
summary, the thermal expectation value of $A$ is approximated by
\begin{eqnarray}
    \langle A\rangle_{\beta}
    &\approx&
    \frac1{\cal Z}\sum_{s}
    \frac{N_{\rm st}^s}{N_{\text R}^s}\sum_{r=1}^{N_{\text R}^s}
    \sum_{j=0}^{M_{\text R}^{s}}
    e^{-\beta\epsilon_{j}^{r}}
    \langle r_{s}|\psi_{j}^{r}\rangle
    \langle\psi_{j}^{r}|A|r_{s}\rangle,
    \label{eqn:lancz:a}
    \\
    {\cal Z}
    &\approx&
    \sum_{s}
    \frac{N_{\rm st}^s}{N_{\text R}^s}\sum_{r=1}^{N_{\text R}^s}
    \sum_{j=0}^{M_{\text R}^{s}}
    e^{-\beta\epsilon_{j}^{r}}
    \left|\langle r_{s}|\psi_{j}^r\rangle\right|^{2},
    \label{eqn:lancz:psi}
\end{eqnarray}
where the index $s$ denotes the summation over all symmetry
sectors of the Hilbert space with dimension $N_{\text{st}}^{s}$.
If the operator $A$ is a conserved quantity with $\left[{\cal
H},A\right]=0$, we can replace $A$ by its quantum numbers
$A_{j}^{r,s}$, and Eq.~(\ref{eqn:lancz:a}) further simplifies to
\begin{equation}
    \langle A\rangle_{\beta}
    \approx
    \frac1{\cal Z}\sum_{s}
    \frac{N_{\rm st}^s}{N_{\text R}^s}\sum_{r=1}^{N_{\text R}^s}
    \sum_{j=0}^{M_{\text R}^{s}}
     e^{-\beta\epsilon_{j}^{r}}
     A_{j}^{r,s}
    \left|\langle r_{s}|\psi_{j}^r\rangle\right|^{2}.
    \label{eqn:as}
\end{equation}
The general definitions of the volume magnetic susceptibility and the
heat capacity are
\begin{equation}
    \chi_{V}
    =
    \mu_{0}\lim_{B_{\text a}\to0}\frac{\partial^{2}F}{\partial B_{\text a}^{2}},
    \quad
    C_{V}
    =
    -\frac{k_{\text B}}\beta
    \left(\beta^{2}\frac{\partial}{\partial_{\beta}}\right)^{2}F,
\end{equation}
where $\vec H=\vec B_{\text a}/\mu_{0}$ is the external magnetic field
defining the $z$ direction and $F=(-1/\beta)\ln{\cal Z}$ is the
canonical free energy.  Written in terms of the expectation values
defined above, we get the cumulants
\begin{eqnarray}
    \chi
    &=&
    \frac{\beta J_{\text c}}N\left(
    \left\langle\Omega_{z}^{2}\right\rangle_{\beta}
    -\left\langle\Omega_{z}\right\rangle_{\beta}^{2}
    \right),
    \label{eqn:lancz:chi}
    \\
    c_{V}
    &=&
    \frac{\beta^{2}}N\left(
    \left\langle{\cal H}^{2}\right\rangle_{\beta}
    -\left\langle{\cal H}\right\rangle_{\beta}^{2}
    \right)
    \label{eqn:lancz:cv}
\end{eqnarray}
for the molar susceptibility and the specific heat of a tile with $N$
sites, respectively.  In our case, both $\Omega_{z}$ and trivially $\cal H$
commute with $\cal H$, allowing us to use Eq.~(\ref{eqn:as}) to
determine the thermal expectation values.  Furthermore spontaneous
magnetic order is absent because our tiles are finite and we do not
include an external magnetic field, therefore we can safely set
$\left\langle\Omega_{z}\right\rangle_{\beta}=0$.

As mentioned before, in the comparison of calculated values and experimental 
results two strategies are possible. One can identify the maximum position and
value of thermodynamic quantities or perform a fit over the whole temperature range
to extract the exchange model parameters. We will discuss both approaches.

\begin{figure*}
    \centering
    \includegraphics[width=.7\textwidth]{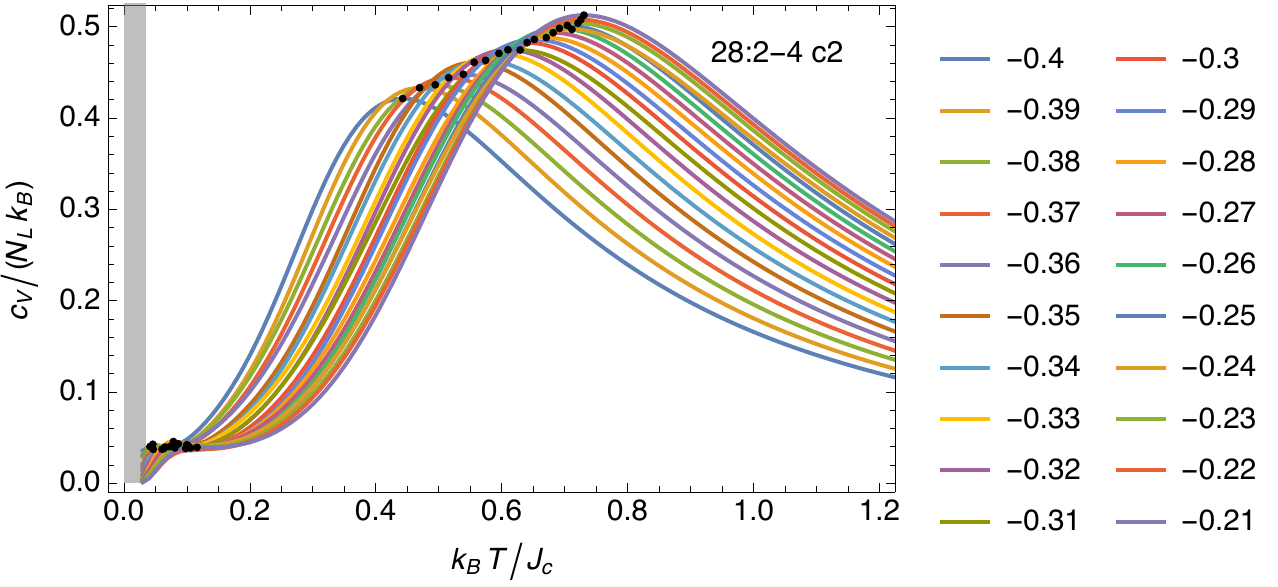}
    \caption{Temperature dependence of the specific heat $c_{V}(T)$ of
    the anisotropic triangular lattice according to
    Eq.~(\ref{eqn:lancz:cv}) for anisotropy parameters $\phi$ in the
    antiferromagnetic phase ranging between $\phi/\pi=-0.4$ and
    $\phi/\pi=-0.21$, see legend.  We used tile 28:2-4 for the
    numerical evaluation of Eq.~(\ref{eqn:lancz:cv}), the grey-shaded
    area at low temperatures illustrates the finite-size gap of order
    ${\cal O}(J_{\text c}/N)$.  The characteristic maxima of
    $c_{V}(T)$ are indicated by the small black dots.}
    \label{fig:cvtaf}
\end{figure*}
\begin{figure*}
    \centering
    \includegraphics[width=.7\textwidth]{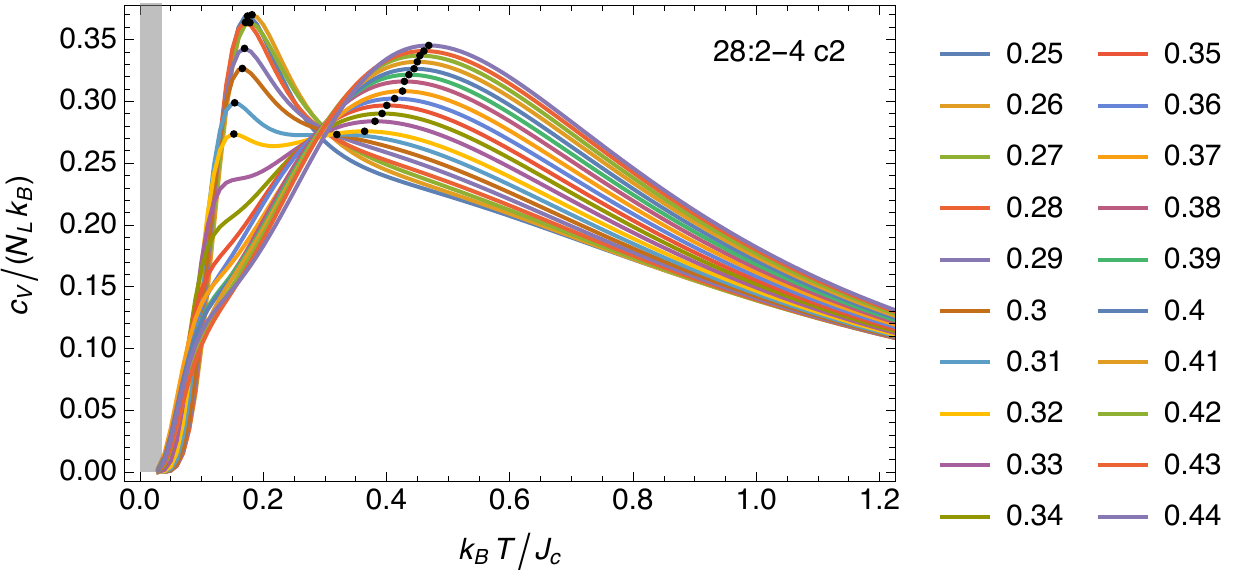}
    \caption{Temperature dependence of the specific heat $c_{V}(T)$ of
    the anisotropic triangular lattice according to
    Eq.~(\ref{eqn:lancz:cv}) for anisotropy parameters $\phi$ in the
    spiral phase ranging between $\phi/\pi=0.25$ (isotropic triangular
    lattice) and $\phi/\pi=0.44$ (crossover to one-dimensional
    chains), see legend.  We used tile 28:2-4 for the numerical
    evaluation of Eq.~(\ref{eqn:lancz:cv}), the grey-shaded area at
    low temperatures illustrates the finite-size gap of order ${\cal
    O}(J_{\text c}/N)$.  The characteristic maxima of $c_{V}(T)$ are
    indicated by the small black dots.}
    \label{fig:cvt}
\end{figure*}
\section{Heat capacity}
\label{sec:heatcapacity}

First we discuss the heat capacity which requires only the cluster
eigenvalues for its evaluation.  A comparison of the theoretical
temperature dependence to experiments is, however, not straightforward
due to the lattice contribution to the heat capacity~\cite{kini:06}.

\subsection{Temperature dependence}
\label{sec:cvt}

To demonstrate characteristic features, Figs.~\ref{fig:cvtaf}
and~\ref{fig:cvt} show the temperature dependence of the specific heat
$c_{V}(T)$ according to Eq.~(\ref{eqn:lancz:cv}) for a selected range
of anisotropy parameters $\phi$ in the antiferromagnetic
(Fig.~\ref{fig:cvtaf}) and the spiral phase (Fig.~\ref{fig:cvt}).  The
FTLM calculations are trustworthy only down to a temperature range of
$T\approx J_{\text c}/(Nk_{\text B})$.  For lower temperatures they
are dominated by the artificial finite size gaps.  This excluded region
is indicated by grey bars in the figures.

For the AF phase with values $-0.4\le\phi/\pi\le-0.21$ a single peak
indicated by dots that shifts continuously to higher values with
increasing $\phi$ is observed.  This is qualitatively clear already
from the eight-site spectrum (Fig.\ref{fig:spectrum}) which shows an
increasing average excitation gap from the ground state in that range
of $\phi$.  Furthermore a plateau (or very flat second maximum) at the
lowest temperatures close to the finite size gap region is visible.

For the spiral phase the values $0.25\le\phi/\pi\le0.44$, 
correspond to the region between the isotropic triangular
lattice with $J_{2}/J_{1}=1$ and deep inside the disordered phase with
quasi-one-dimensional chains ($J_{2}/J_{1}\approx5$).  The tile 28:2-4
was used for the numerical calculations in both cases.

In Fig.~\ref{fig:cvt}, the characteristic maxima of $c_{V}(T)$,
indicated by the small black dots in the figure, have positions which
fall into two different temperature ranges: For the isotropic
triangular case with $\phi/\pi=0.25$, the maximum sits at
$T_{\triangle}\approx0.18J_{\text c}/k_{\text B}$, reducing in
magnitude and shifting slightly to lower temperatures
$T_{\triangle}\approx0.15J_{\text c}/k_{\text B}$ upon increasing
$\phi$ from its isotropic value to $\phi/\pi\approx0.31$ or
$J_{2}/J_{1}\approx1.5$.  At this point, a second maximum develops
starting at $T_{\parallel}\approx0.32J_{\text c}/k_{\text B}$,
shifting to higher temperatures as $\phi$ is increased towards the
disordered region.  Furthermore, the maximum $T_{\triangle}$
characteristic for the triangular lattice rapidly disappears with
increasing $\phi$.  Only the $C_{V}(T)$ curves for $\phi/\pi=0.31$ and
$\phi/\pi=0.32$ show both maxima simultaneously.  At high temperatures
$T\gg J_{\text c}/k_{\text B}$, all the curves show the $(1/T)^{2}$
temperature dependence expected from Eq.~(\ref{eqn:cvht}).

\subsection{Exchange anisotropy dependence of peak position and value}

\begin{figure}
    \centering
    \includegraphics[width=\columnwidth]{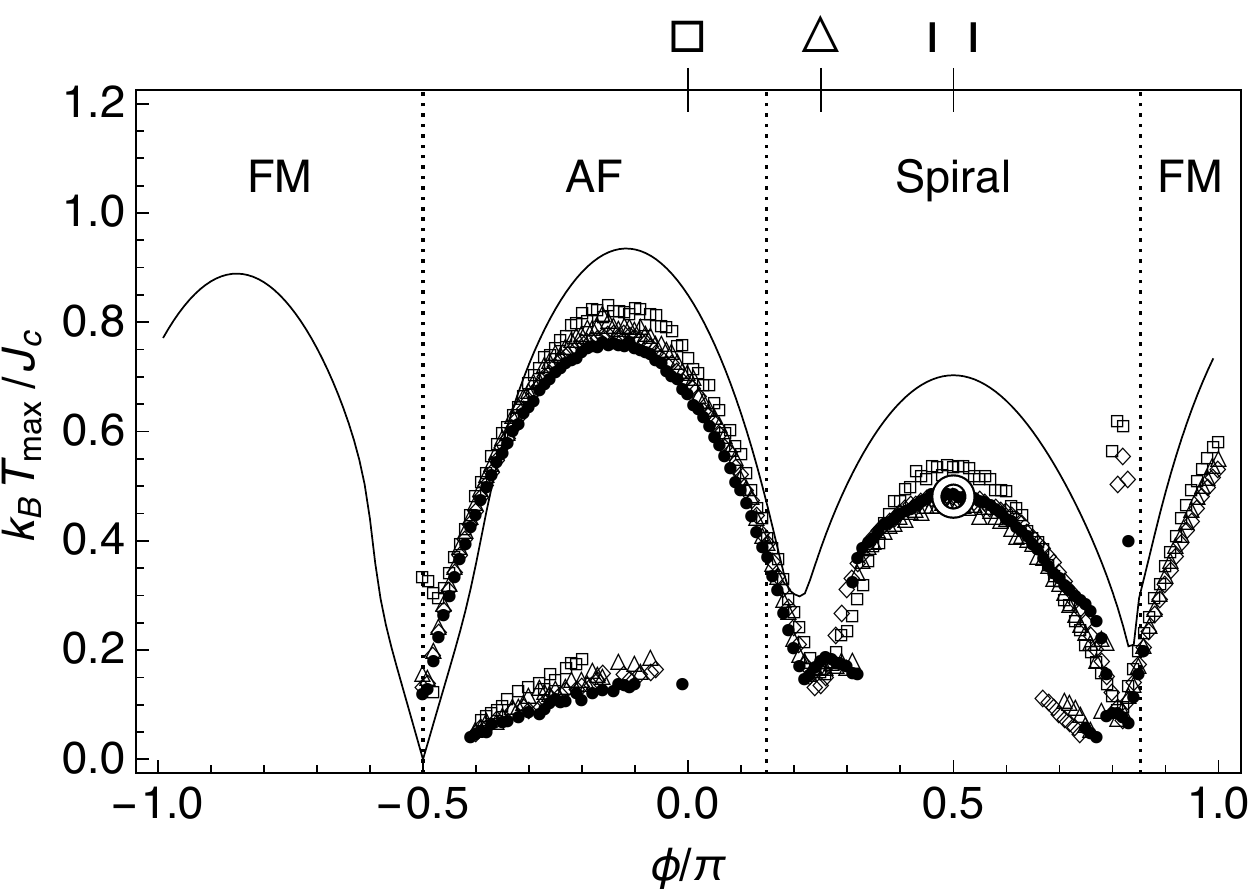}
    \includegraphics[width=\columnwidth]{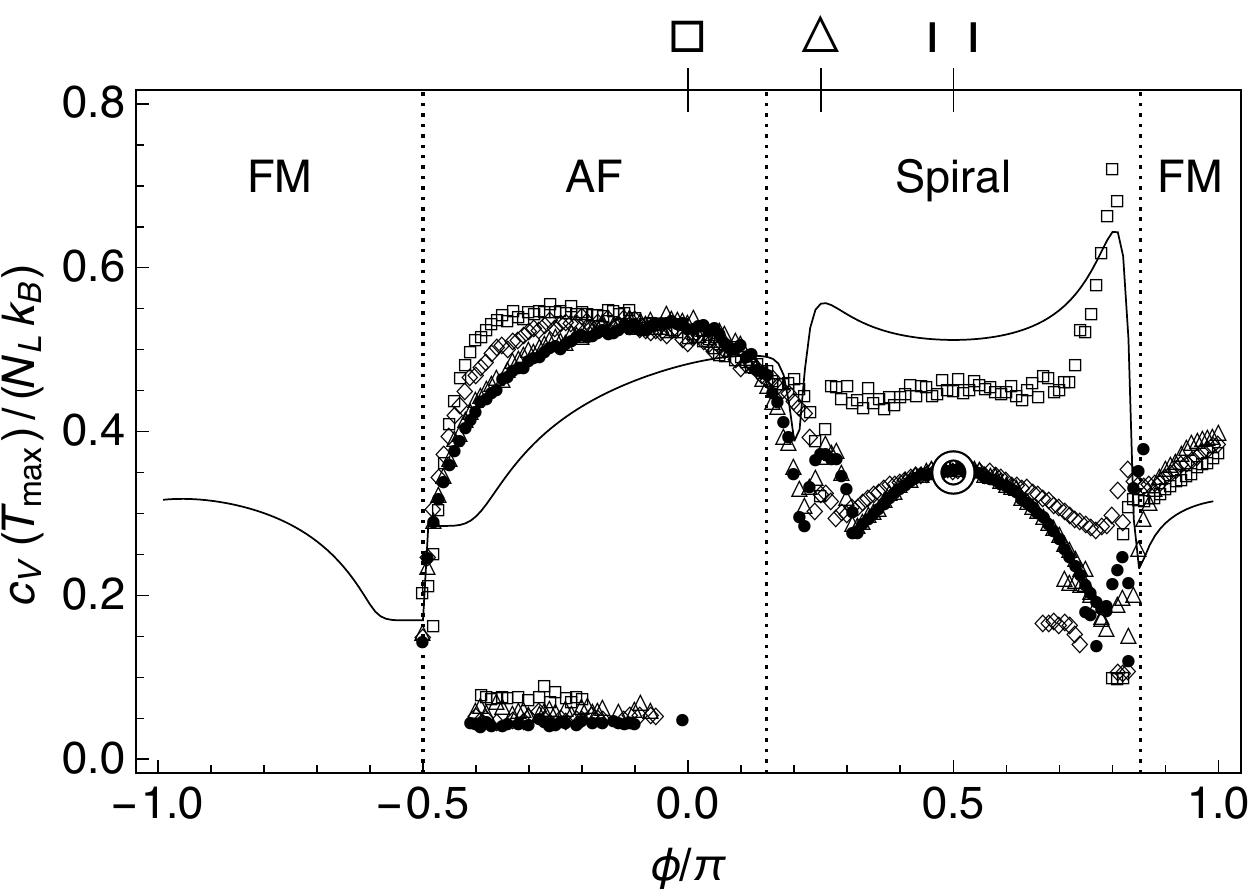}
    \caption{Maximum positions (top) and values (bottom) of the
    specific heat $c_{V}(T)$ as a function of the anisotropy parameter
    $\phi$.  Solid lines: exact solution for tile 8:2-2, squares: tile
    16:4-0, triangles: tile 20:2-4, diamonds: tile 24:4-0, filled
    circles: tile 28:2-4.  The white rings mark the corresponding exact
    values for the one-dimensional chain~\cite{bonner:64}.}
    \label{fig:cvmax}
\end{figure}
Fig.~\ref{fig:cvmax} shows a compilation of the maximum positions and
values of $c_{V}(T)$ for the five different tiles we have used as a
function of the anisotropy parameter $\phi$.  The solid lines denote
the exact solution for the tile 8:2-2, see Eq.~(\ref{eqn:cveight}) and
Table~\ref{tbl:spectrum}.  The black dots denote the maxima for the
tile 16:4-0, the open triangles for the tile 20:2-4, the solid
diamonds for the tile 24:4-0, the open squares for the tile 28:2-4.
Furthermore, the exact results taken from Ref.~\cite{bonner:64} for the
one-dimensional chain, $c_{V}(T_{\text{max}})/(N_{\text L}k_{\text
B})=0.35$ and $k_{\text B}T_{\text{max}}/J_{\text c}=0.481$, are
marked with the white circles and serve as a gauge to judge how
close a cluster of given size approaches the thermodynamic limit.

The eight-site cluster, introduced as an illustration of the model and
its overall spectrum, clearly is not very useful quantitatively to discuss the heat
capacity in the thermodynamic limit.  Only the overall behavior of the
maximum temperature is qualitatively similar to our findings for the
larger tilings, however the two minima in the spiral phase are shifted
towards the classical phase borders.  The values
$c_{V}(T_{\text{max}})$  for the eight-site cluster are monotonically increasing in the
antiferromagnetic phase, followed by a minimum at $\phi/\pi\approx0.1$
and a maximum at the isotropic point, $\phi/\pi=0.25$.  On the
ferromagnetic-$J_{1}$ side for $1/2<\phi/\pi\le1$,
$c_{V}(T_{\text{max}})$ increases again to a second maximum, followed
by a minimum at the crossover to the ferromagnet at
$J_{2}/|J_{1}|=1/2$. At the one-dimensional point $J_{1}=0$, naturally both 
$T_{\text{max}}$ and $c_{V}(T_{\text{max}})$ differ strongly from the 
exact values.

Also the 16-site tiling shows similar features.  For the larger
clusters of size $N=20$ and more, both maximum temperature positions
and maximum values of the specific heat clearly show a double-peak
structure as function of $\phi$.  The positions $T_{\text{max}}$
decrease with increasing cluster size, apart from the regions around
the isotropic point and in the spiral phase near the ferromagnet.  At
the one-dimensional point, agreement with the infinite-chain
result~\cite{bonner:64} (white circles) is achieved already with
$N=20$ tiling.  And therefore results for the $N=28$ tile represent
well the thermodynamic limit as far as peak position and height is
concerned.  Denoting the largest value in each sector of
Fig.~\ref{fig:cvmax} by $T^*_{\text{max}}$ we observe that
$T^*_{\text{max}}(\text{AF})/T^*_{\text{max}}(\text{spiral})=1.58$.
This asymmetry in $T_{\text{max}}(\phi)$ is considerably larger than
the corresponding one in the square lattice $J_1$-$J_2$ model where
$T^*_{\text{max}}(\text{AF})/T^*_{\text{max}}(\text{CAF})=1.26$.  (The
columnar AF (CAF) phase replaces the spiral phase in this model.)
Similarly for the asymmetry ratios of peak values in the triangular
case we get
$c_V[T^*_{\text{max}}](AF)/c_V[T^*_{\text{max}}](\text{spiral})=1.51$
much larger than
$c_V[T^*_{\text{max}}](AF)/c_V[T^*_{\text{max}}](\text{CAF})=1.10$ in
the square lattice.  Inside the AF and spiral sectors the lowest
$T_{\text{max}}$ is reached close to the isotropic triangular point with
$T^{\triangle}_{\text{max}}\approx0.18J_{\text c}/k_{\text B}$ because there
exchange frustration is most pronounced (Fig.~\ref{fig:frustratio}).  This leads to a high density
of low energy excitations (c.f. Fig.~\ref{fig:spectrum}) and therefore a low
$T^{\triangle}_{\text{max}}$.

Furthermore, in parts of the antiferromagnetic as well as the spiral
phase, a second maximum $c_{V}(T_{\text{max}2})$ at very low
temperatures appears.  $T_{\text{max}2}$ depends roughly linearly on
$\phi$ while $c_{V}\left(T_{\text{max}2}\right)$ remains constant and
small, see also Fig.~\ref{fig:cvtaf}.  With the 28-site tiling being
the largest possible, we cannot judge whether this second-low
temperature maximum in the antiferromagnetic phase merely is a
finite-size effect.  This is different from the situation in the
spiral phase, where the partially observed second maximum is at
temperatures $T_{\text{max}2}$ far larger than the finite-size gap,
see also Fig.~\ref{fig:cvt}.

Similar to the behavior around the isotropic point, at the crossover
to the ferromagnetic region a second maximum gradually appears in
$c_{V}(T)$ which then evolves to the ``ferromagnetic'' maximum while
the ``spiral'' maximum disappears.  This also happens at comparatively
low temperatures with an irregular behavior of the maximum values as
function of $\phi$.  We attribute this irregularity to the fact that
in particular near the borders of the spiral phase the ground state
and possible low-lying excited states have incommensurate ordering
vectors $\vec Q=\vec Q_{\text{AF}}+\delta\vec Q$ and $\vec
Q_{\text{FM}}+\delta\vec Q$ respectively with $|\delta\vec Q|\ll1$.
These are not contained in the coarse grid of crystal momenta of our finite
tiles.

\section{Susceptibility}
\label{sec:susceptibility}

\begin{figure}
    \centering
    \includegraphics[width=\columnwidth]{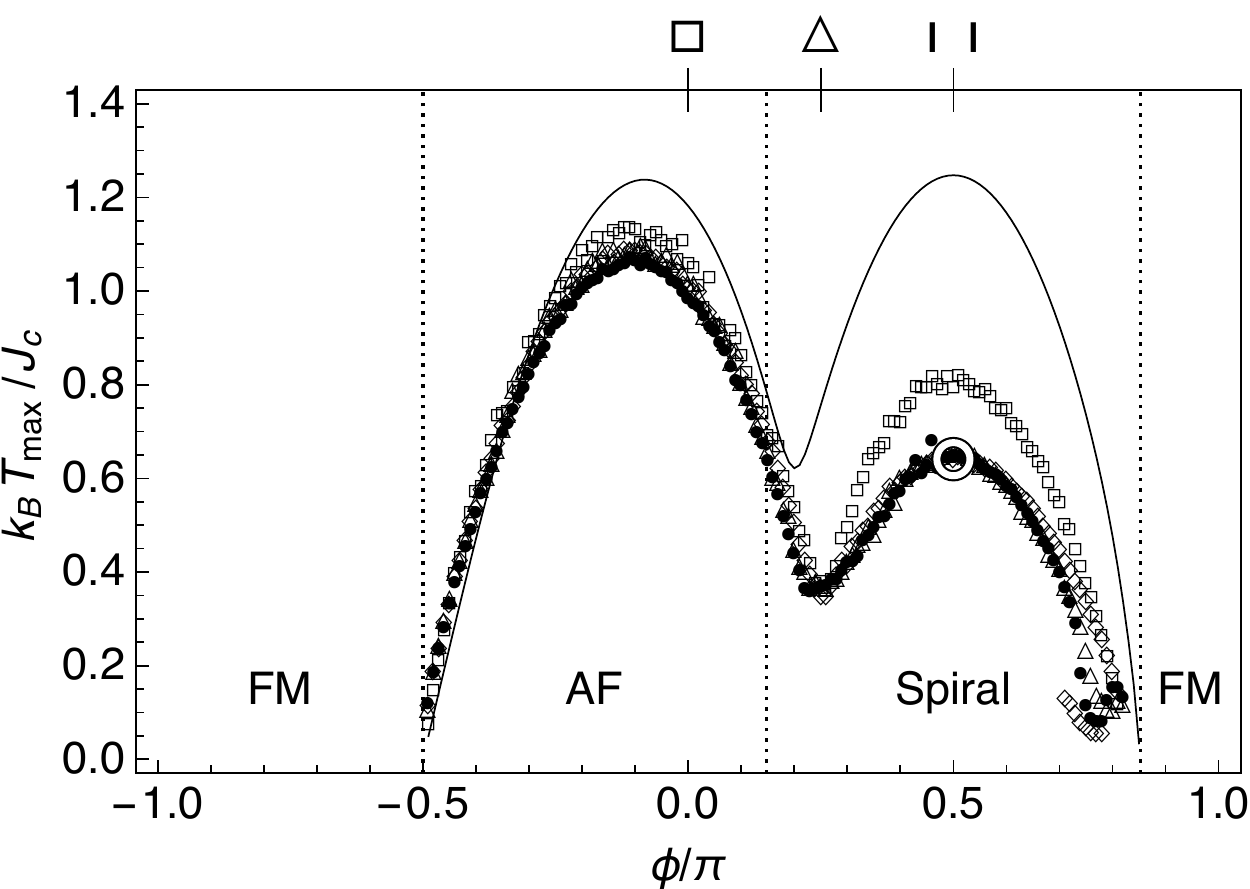}
    \includegraphics[width=\columnwidth]{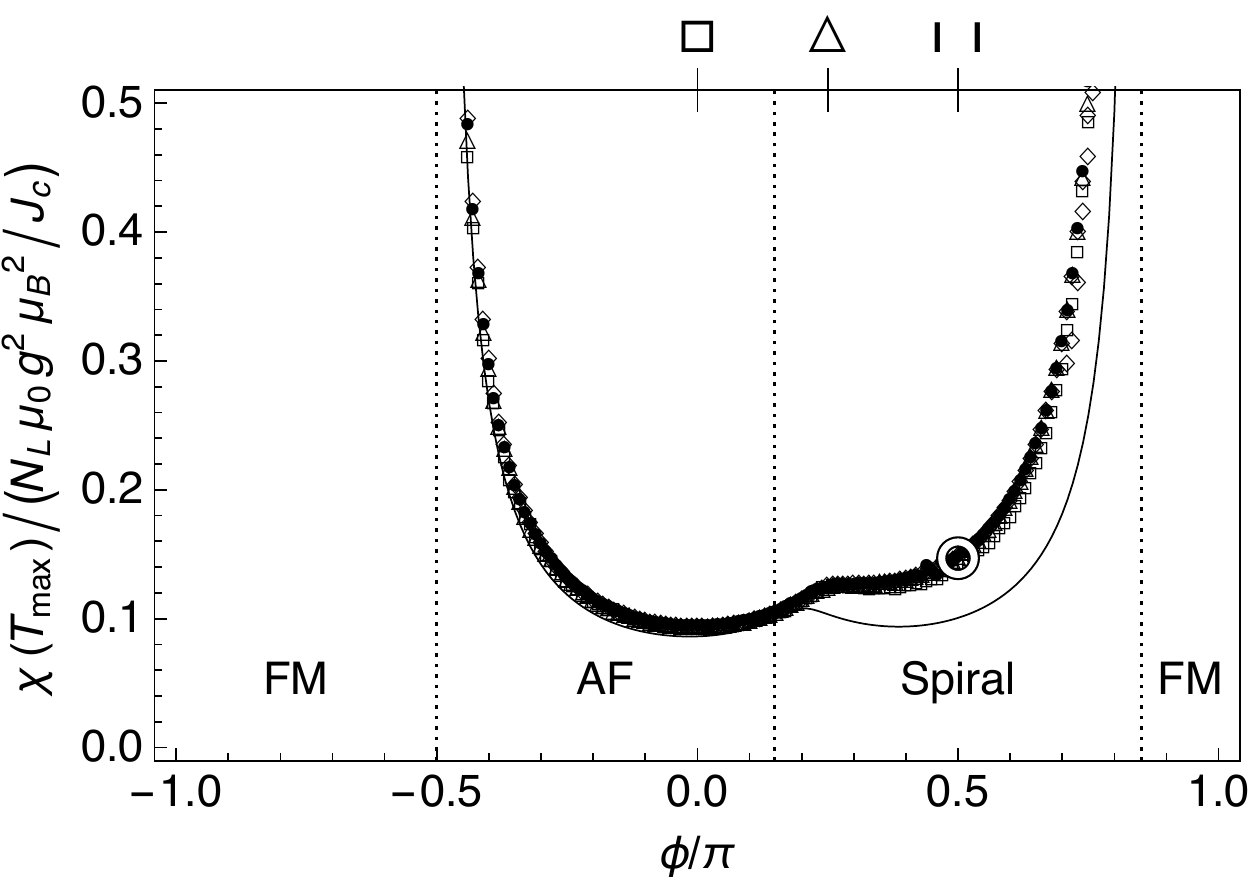}
    \caption{Maximum positions (top) and values (bottom) of the static
    susceptibility $\chi(T)$ as a function of the anisotropy parameter
    $\phi$.  Solid lines: exact solution for tile 8:2-2, squares: tile
    16:4-0, triangles: tile 20:2-4, diamonds: tile 24:4-0, filled
    circles: tile 28:2-4.  The white rings mark the corresponding exact
    values for the one-dimensional chain~\cite{bonner:64}.}
    \label{fig:chimax}
\end{figure}
\begin{figure*}
    \centering
    \includegraphics[width=.8\textwidth]{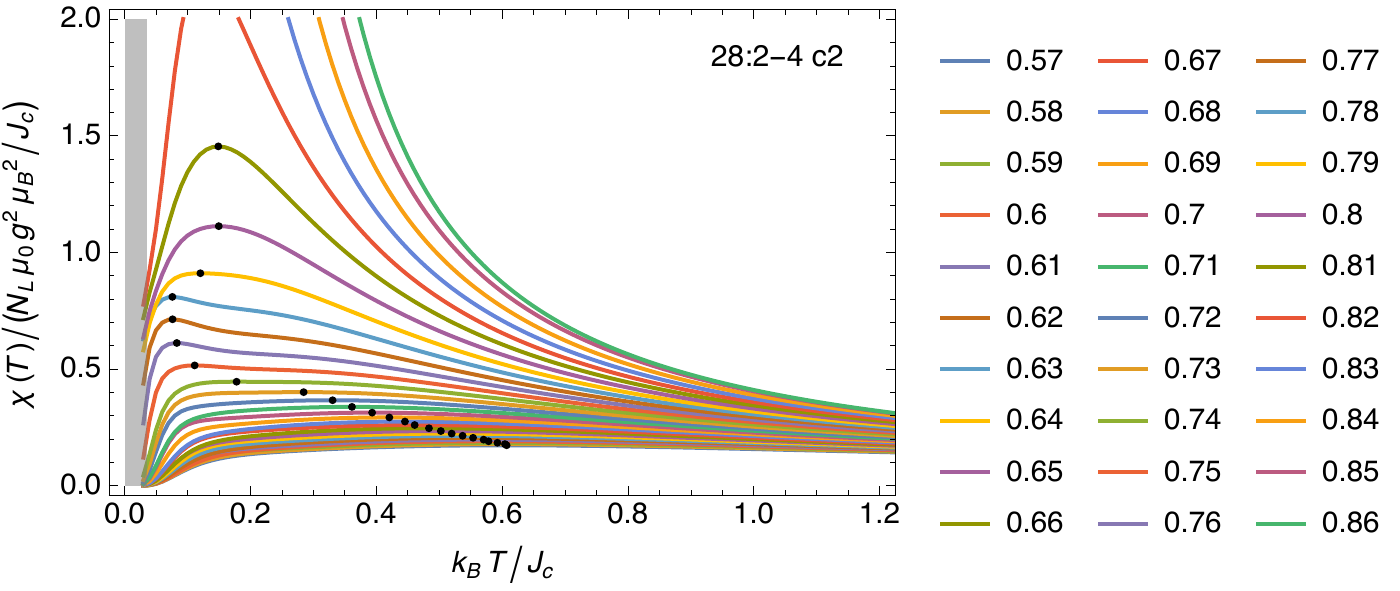}
     \caption{Temperature dependence of the static susceptibility
     $\chi(T)$ of the anisotropic triangular lattice according to
     Eq.~(\ref{eqn:lancz:chi}) for anisotropy parameters $\phi$ in the
     spiral phase ranging between $\phi/\pi=0.57$ and $\phi/\pi=0.86$,
     see legend.  We used tile 28:2-4 for the numerical evaluation of
     Eq.~(\ref{eqn:lancz:chi}), the gray-shaded area at low
     temperatures illustrates the finite-size gap of order ${\cal
     O}(J_{\text c}/N)$.  The characteristic maxima of $\chi(T)$ are
     indicated by the small black dots.}
   \label{fig:chit}
\end{figure*}
Evaluation of the susceptibility with FTLM is more involved because
it requires the calculation of magnetic moment matrix elements. On the
other hand this quantity can be more easily compared to experimental 
results. Most known exchange parameters for spin systems are due to 
the application of this method.

\subsection{Temperature dependence}
\label{sec:chit}

Similar to the specific heat, we have calculated the temperature
dependence of the static magnetic susceptibility $\chi(T)$ according to
Eq.~(\ref{eqn:lancz:chi}) with the tilings shown in
Fig.~\ref{fig:tiles}.  Fig.~\ref{fig:chit} shows the results in a
selected range of the anisotropy parameter, $0.57\le\phi/\pi\le0.86$,
calculated with tiling 28:2-4.  This parameter range corresponds to
the interpolation between ferromagnetically coupled ($J_{1}<0$)
quasi-one-dimensional antiferromagnetic chains with
$J_{2}/J_{1}\approx-4.5$ and the crossover to the ferromagnet at the
classical boundary $J_{2}/J_{1}=-1/2$.  For the former, a very broad
maximum is characteristic which gradually evolves into the $T=0$
divergence of $\chi(T)$ in the ferromagnetic phase, which is the 
expected behavior.

\subsection{Exchange anisotropy dependence}

In the same way as for the specific heat, we follow the positions and
values of the characteristic maxima of $\chi(T)$ as a function of the
anisotropy parameter, using the tiles displayed in
Fig.~\ref{fig:tiles} again.  This is shown in Fig.~\ref{fig:chimax}.
The solid lines denote the eight-site results according to
Eq.~(\ref{eqn:chieight}).  For the same reasons as discussed for the
heat capacity, strong deviations from the larger tilings occur in
particular in the spiral phase.  In contrast to $c_{V}(T)$, the
maximum positions and values of $\chi(T)$ are already converged for
tilings of size $N=20$ and larger, apart from the crossover to the
ferromagnetic phase, where a tile-dependence of the maximum
temperatures clearly can be observed.

Similar as for the specific heat the $T_{\text{max}}(\phi)$ dependence
has a pronounced asymmetry in AF and spiral sectors.  Denoting again
the largest value in each sector by $T^*_{\text{max}}$ we obtain for
the triangular lattice
$T^*_{\text{max}}(\text{AF})/T^*_{\text{max}}(\text{spiral})=1.58$
which is much larger than the asymmetry value
$T^*_{\text{max}}(\text{AF})/T^*_{\text{max}}(\text{CAF})=1.11$ for
the square lattice case.  Similar to specific heat behavior the
$T_{\text{max}}(\phi)$ minimum inside AF and spiral sectors is reached
around the most frustrated isotropic triangular position
$T^{\triangle}_{\text{max}}\approx0.35J_{\text c}/k_{\text B}$.

At the one-dimensional point, as for the heat capacity,
the results from Ref.~\cite{bonner:64}, $k_{\text
B}T_{\text{max}}/J_{\text c}=0.641$ and
$\chi(T_{\text{max}})/(N_{\text L}\mu_{0}(g\mu_{\text B})^{2}/J_{\text
c})=0.147$ are accurately reproduced with our method.  In general we
think that in particular our results on the magnetic susceptibility
can be used to accurately determine both exchange constants $J_{1}$
and $J_{2}$ individually, which we show in the following section.

\section{Application to C\lowercase{s}$_{\text
2}$C\lowercase{u}C\lowercase{l}$_{\text 4}$ and C\lowercase{s}$_{\text
2}$C\lowercase{u}B\lowercase{r}$_{\text 4}$}
\label{CsCu}
\begin{figure}
    \centering
    \includegraphics[width=\columnwidth]{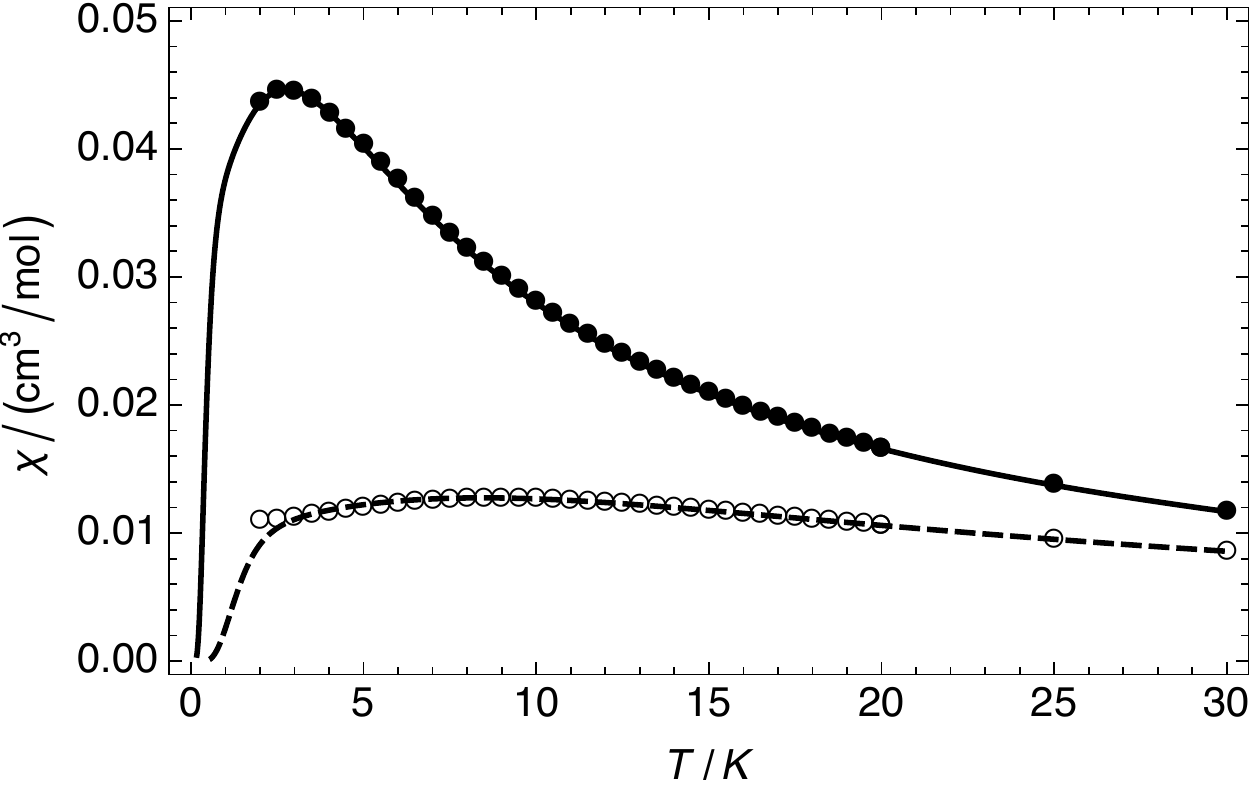}
    \includegraphics[width=\columnwidth]{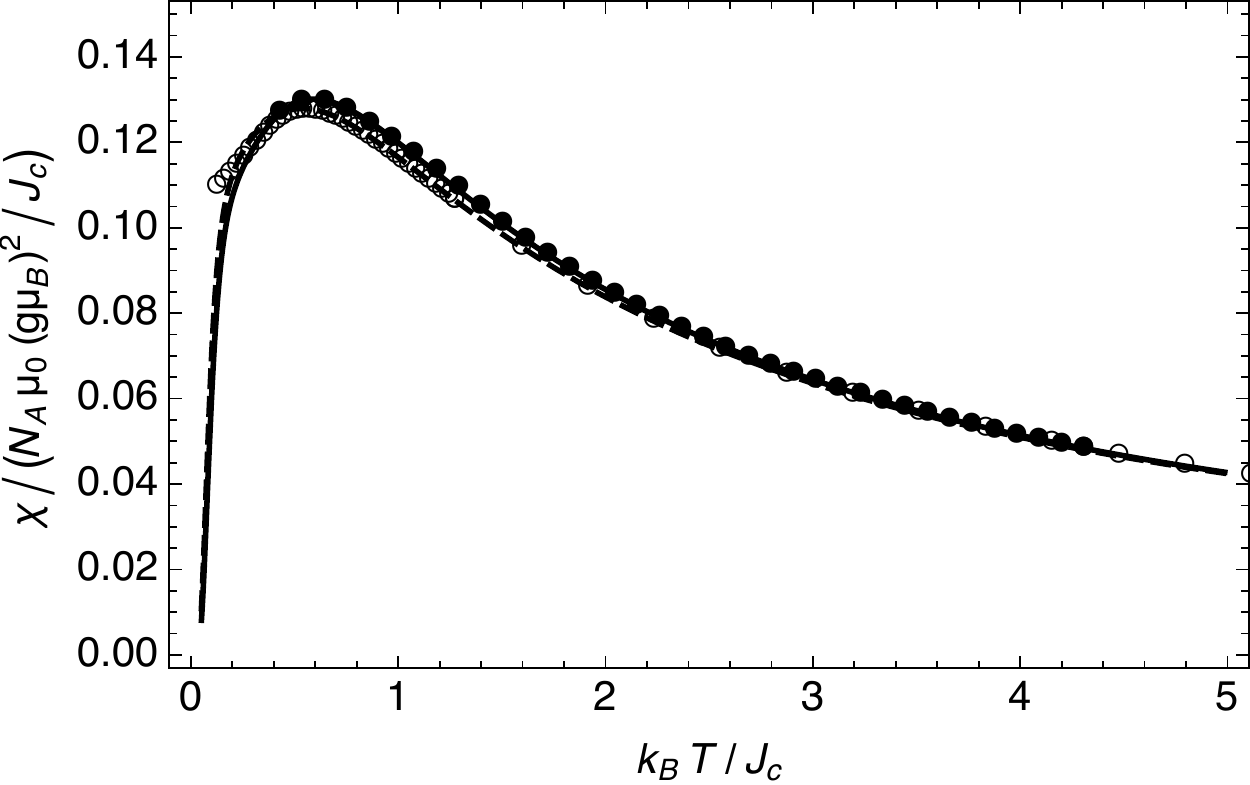}
    \caption{Temperature dependence of the magnetic susceptibility
    $\chi(T)$ of Cs$_{\text 2}$CuCl$_{\text 4}$ (dots and solid line)
    and Cs$_{\text 2}$CuBr$_{\text 4}$ (open circles and dashed line).
    The top plot displays the experimental data (symbols, taken from
    Ref.~\cite{cong:11}) together with the fits of our FTLM data
    (lines, fitted values see Table~\ref{tbl:compound}).  The bottom
    plot displays exactly the same data, this time in dimensionless
    units using $J_{\text c}$, $\phi$, and $g$ for the two compounds
    from Table~\ref{tbl:compound}.}
    \label{fig:chicscucl}
\end{figure}
\begin{table*}
    \caption{Comparison of exchange parameters for Cs$_{\text
    2}$CuCl$_{\text 4}$ and Cs$_{\text 2}$CuBr$_{\text 4}$ as
    determined from thermodynamic FTLM-fit and direct spectroscopic
    INS (Cs$_{\text 2}$CuCl$_{\text 4}$ only) and ESR methods at
    $H>H_{\text{sat}}$.  Exchange constant $J_2$ corresponds to the
    crystallographic $b$ direction and $J_1$ to the zigzag bonds in
    the $bc$ plane.}
    \begin{ruledtabular}
        \begin{tabular}{ccccccccc}
	    compound 
	    &
            method
            &
            $J_1/\text{meV}$
            &
            $J_{2}/\text{meV}$
            &
            $J_{\text c}/\text{meV}$
            &
            $J_{2}/J_{1}$
            &
            $\phi/\pi$
            &
            $g$
            &
            Ref.
            \\
            \hline
	    Cs$_{\text 2}$CuCl$_{\text 4}$
	    &
            FTLM
            &
            $0.11$
            &
            $0.38$
            &
            $0.40$
            &
            $3.45$
            &
            $0.41$
            &
            $2.06$
            &
            this work
            \\
	    &
            INS
            &
            $0.128$
            &
            $0.374$
            &
            $0.40$
            &
            $2.92$
            &
            $0.40$
            &
            $2.19$
            &
            \cite{coldea:02}
            \\
	    &
            ESR
            &
            $0.12$
            &
            $0.41$
            &
            $0.43$
            &
            $3.42$
            &
            $0.41$
            &
            $2.08$
            &
            \cite{zvyagin:14}
	    \\
	    \hline
	    Cs$_{\text 2}$CuBr$_{\text 4}$
	    &
	    FTLM
	    &
	    $0.5$
	    &
	    $1.26$
	    &
	    $1.35$
	    &
	    $2.52$
	    &
	    $0.38$
	    &
	    $2.04$
	    &
	    this work
	    \\
	    &
	    ESR
	    &
	    $0.53$
	    &
	    $1.28$
	    &
	    $1.38$
	    &
	    $2.44$
	    &
	    $0.38$
	    &
	    $2.09$
	    &
	    \cite{zvyagin:14}
	\end{tabular}
    \end{ruledtabular}
      \protect\label{tbl:compound}
\end{table*}
As a demonstration of the usefulness of the method, we apply our
findings to the compounds Cs$_{\text 2}$CuCl$_{\text 4}$ and
Cs$_{\text 2}$CuBr$_{\text 4}$.  Fig.~\ref{fig:chicscucl} shows the
temperature dependence of the magnetic susceptibility.  The dots and
the open circles denote the experimental data taken from
Ref.~\cite{cong:11}, the solid and dashed lines denote the
corresponding fits with our FTLM data.  In order to avoid
misunderstandings, we have plotted the data and the curves in two
different ways: The top plot in Fig.~\ref{fig:chicscucl} displays
$\chi(T)$ in electromagnetic (CGS) units from where we have determined
the values for our model parameters reproduced in
Table~\ref{tbl:compound}.  Using these fitted values for $J_{\text
c}$, $\phi$, and $g$, we have plotted the same data again in
dimensionless units in the bottom plot of Fig.~\ref{fig:chicscucl}.
The main effect of replacing Cl with Br is an increase of the overall
energy scale $J_{\text c}$ by a factor $3.4$, whereas the anisotropy
angle $\phi$ changes only by about $5\,\%$.  Thus only this large
change in $J_{\text c}$ is responsible for the decrease of the maximum
in $\chi(T)$, the broadening of the maximum, and the shift of the
maximum towards higher temperatures.  The anisotropy ratio and
therefore the position of Cs$_{\text 2}$CuBr$_{\text 4}$ in the phase
diagram remains essentially the same as for Cs$_{\text 2}$CuCl$_{\text
4}$, it is only slightly moved away from the quasi-one-dimensional 
regime.  Historically the first measurement of $\chi(T)$ of Cs$_{\text
2}$CuCl$_{\text 4}$ yielded slightly different
values~\cite{carlin:85}, however the data analysis was performed over
a limited temperature range assuming weakly coupled one-dimensional
chains.

As pointed out in Sec.~\ref{sec:cvt}, the steep decrease of $\chi(T)$
for $T\to0$ in both cases is an artifact of the finiteness of the
tiling used for the FTLM calculations.  Assuming a finite-size gap
$\Delta\approx J_{\text c}/N$, the corresponding temperatures where we
expect finite-size effects to dominate are
$T_{\Delta}\approx0.2\,\text K$ for Cs$_{\text 2}$CuCl$_{\text 4}$ and
$T_{\Delta}\approx0.6\,\text K$ for Cs$_{\text 2}$CuBr$_{\text 4}$.
Experimentally, the lowest temperature was
$T_{\text{min}}\approx2\,\text K$, well above the finite-size gaps.

Our result from the FTLM fit to thermodynamic data are compared in
Table~\ref{tbl:compound} to results from direct spectroscopic methods:
Inelastic neutron scattering (INS)~\cite{coldea:02} (Cs$_{\text
2}$CuCl$_{\text 4}$ only) and electron spin resonance
(ESR)~\cite{zvyagin:14}, both in fields above the saturation 
field~\cite{schmidt:14}
\begin{equation}
    \mu_{0}H_{\text{sat}}=\frac{2S}{g\mu_{\text B}}
    \frac{(J_{1}+2J_{2})^{2}}{2J_{2}},
\end{equation}
which is $\mu_{0}H_{\text{sat}}\approx8.4\,\text T$ for Cs$_{\text
2}$CuCl$_{\text 4}$ and $\mu_{0}H_{\text{sat}}\approx30\,\text T$ for
Cs$_{\text 2}$CuBr$_{\text 4}$.  Taking the spectroscopic data at high
fields has the advantage that the ground state is fully polarized,
corresponding to a single ``all spins up'' spinor product state
$|\text{FM}\rangle$.  The excitations on top of this are the $N$
orthonormal single-particle excitations
$|\psi_{i}\rangle=(1/\sqrt{2S})S_{i}^{-}|\text{FM}\rangle$, $i=1\ldots
N$ with $N={\cal O}(N_{\text L})$.  Because $\left[{\cal
H},\Omega_{z}\right]=0$, the Hamiltonian can not generate more than
these one-spin-flip states, and its spectrum can be determined exactly
by Fourier transform.

The agreement between the different methods is almost perfect, for
exchange parameters as well as g-factors.  As already mentioned in
Ref.~\cite{schmidt:14} if Cs$_{\text 2}$CuCl$_{\text 4}$ is
interpreted as a purely 2D system this set of exchange parameters puts
the compound very close to the quasi-1D spin liquid regime centered at
$\phi=0.5\pi$.  In reality, however, the magnetic order is stabilized
below $T_{\text N} = 0.62$ K by a finite inter-plane coupling of the
order $J_\perp\approx0.017\,\text{meV}$ along the crystallographic $a$
direction~\cite{coldea:03}.

For Cs$_{\text 2}$CuBr$_{\text 4}$, we see deviations of the
experimental data from our result at the lowest temperatures.  These
are not due to impurities~\cite{cong:11,cong:14}, but can be regarded
as an indication for a tendency towards magnetic order in this
compound.  The overall energy scale $J_{\text c}$ is more than three
times larger than for Cs$_{\text 2}$CuCl$_{\text 4}$, however the
estimate for the anisotropy angle $\phi$ is essentially the same.  We
therefore expect that Cs$_{\text 2}$CuBr$_{\text 4}$ orders
magnetically as well, possibly at a higher temperature than its Cl
counterpart.

The agreement of our thermodynamic analysis with direct spectroscopic
results gives us confidence that the FTL method may be applied to the
analysis of the whole Cs$_{\text 2}$CuCl$_{\text 4-x}$Br$_{\text x}$
substitutional series~\cite{cong:11,cong:14}.

\section{Conclusion and Outlook}
\label{conclusion}

In this work we have shown that finite temperature Lanczos method for
finite clusters is a versatile tool to investigate the little known
finite temperature properties of frustrated triangular quantum
magnets, in contrast to QMC approach which is not suitable in this
case.  The FTL method complements the analytical spin wave approach
which is useful only for the very low temperature regime and is a more
straightforward alternative to the high temperate series expansion
method.

For this quantum spin model one can use a single control parameter and
tune the system through the phase diagram where several special cases
with very different frustration degree like frustrated N\'eel,
unfrustrated HAF, frustrated spiral and isotropic triangular
$120^\circ$ phase as well as unfrustrated spin chains and frustrated
FM may be realized.

For the largest investigated cluster with $N=28$ sites we obtain a
trustworthy representation of the thermodynamic limit behavior since
the exact known result for the spin chain case is produced very well
and shows little difference to the $N=24$ cluster.  As a main result
we gave the systematic variation of peak position and peak height of
specific heat and susceptibility as function of $J_2/J_1$.  Both
values are strongly suppressed for the most frustrated isotropic
triangular magnet (Figs.~\ref{fig:cvmax} and~\ref{fig:chimax}).
Furthermore a surprisingly large asymmetry in particular for the
maximum position exists between the AF and spiral phase region.  It is
considerably larger than between the AF and CAF regions in the square
lattice $J_1$-$J_2$ model. In addition we found that the simple single
peak shape of $c_V(T)$ and $\chi(T)$ may be distorted due to smaller
side maxima or shoulders at lower temperature for some ranges of the
control parameter.

The most reliable method to extract the exchange parameters is the
fitting of $\chi(T)$ over the whole temperature range (above the
finite size gap) using FTLM results.  We have demonstrated this for
two of the most typical 2D triangular magnets, Cs$_{\text
2}$CuCl$_{\text 4}$ and Cs$_{\text 2}$CuBr$_{\text 4}$.  We have
obtained excellent agreement with results from the INS investigation
and with ESR results in the fully polarized state.  Our method has the
additional advantage that it can easily be extended to the whole
substitutional series Cs$_{\text 2}$CuCl$_{\text 4-x}$Br$_{\text x}$
($0\leq x \leq4$)~\cite{cong:11} to extract the systematic variation
of frustration anisotropy control parameter $\phi$ and energy scale
$J_{\text c}$ as a function of Br concentration~\cite{schmidt:15b}.
We note that our method can also be applied to magnetocaloric
measurements of the organic charge-transfer salts where localized
$S=1/2$ magnetic moments are residing on a possibly distorted
triangular lattice as well.  This includes the
Et$_{x}$Me$_{4-x}$Z$[$Pd$($dmit$)_{2}]_{2}$ ($x=0,1,2$) family of
compounds~\cite{itou:08,tamura:06} as well as
$\kappa$-$($ET$)_{2}$B$($CN$)_{4}$ and
$\kappa$-$($ET$)_{2}$Cu$_{2}($CN$)_{3}$~\cite{yoshida:15} and
$\kappa$-$($BEDT-TTF$)_{2}$Cu$[$N$($CN$)_{2}]$Cl~\cite{hamad:05}.

\begin{acknowledgments}
    We thank P.~Cong, B.~Wolf, and M.~Lang for generously supplying us
    their experimental data discussed here.
\end{acknowledgments}

\bibliography{ftlm-triangle}

\begin{thebibliography}{37}%
\makeatletter
\providecommand \@ifxundefined [1]{%
 \@ifx{#1\undefined}
}%
\providecommand \@ifnum [1]{%
 \ifnum #1\expandafter \@firstoftwo
 \else \expandafter \@secondoftwo
 \fi
}%
\providecommand \@ifx [1]{%
 \ifx #1\expandafter \@firstoftwo
 \else \expandafter \@secondoftwo
 \fi
}%
\providecommand \natexlab [1]{#1}%
\providecommand \enquote  [1]{``#1''}%
\providecommand \bibnamefont  [1]{#1}%
\providecommand \bibfnamefont [1]{#1}%
\providecommand \citenamefont [1]{#1}%
\providecommand \href@noop [0]{\@secondoftwo}%
\providecommand \href [0]{\begingroup \@sanitize@url \@href}%
\providecommand \@href[1]{\@@startlink{#1}\@@href}%
\providecommand \@@href[1]{\endgroup#1\@@endlink}%
\providecommand \@sanitize@url [0]{\catcode `\\12\catcode `\$12\catcode
  `\&12\catcode `\#12\catcode `\^12\catcode `\_12\catcode `\%12\relax}%
\providecommand \@@startlink[1]{}%
\providecommand \@@endlink[0]{}%
\providecommand \url  [0]{\begingroup\@sanitize@url \@url }%
\providecommand \@url [1]{\endgroup\@href {#1}{\urlprefix }}%
\providecommand \urlprefix  [0]{URL }%
\providecommand \Eprint [0]{\href }%
\providecommand \doibase [0]{http://dx.doi.org/}%
\providecommand \selectlanguage [0]{\@gobble}%
\providecommand \bibinfo  [0]{\@secondoftwo}%
\providecommand \bibfield  [0]{\@secondoftwo}%
\providecommand \translation [1]{[#1]}%
\providecommand \BibitemOpen [0]{}%
\providecommand \bibitemStop [0]{}%
\providecommand \bibitemNoStop [0]{.\EOS\space}%
\providecommand \EOS [0]{\spacefactor3000\relax}%
\providecommand \BibitemShut  [1]{\csname bibitem#1\endcsname}%
\let\auto@bib@innerbib\@empty
\bibitem [{\citenamefont {Nakatsuji}\ \emph {et~al.}(2010)\citenamefont
  {Nakatsuji}, \citenamefont {Nambu},\ and\ \citenamefont
  {Onoda}}]{nakatsuji:10}%
  \BibitemOpen
  \bibfield  {author} {\bibinfo {author} {\bibfnamefont {S.}~\bibnamefont
  {Nakatsuji}}, \bibinfo {author} {\bibfnamefont {Y.}~\bibnamefont {Nambu}}, \
  and\ \bibinfo {author} {\bibfnamefont {S.}~\bibnamefont {Onoda}},\ }\href
  {\doibase 10.1143/JPSJ.79.011003} {\bibfield  {journal} {\bibinfo  {journal}
  {J. Phys. Soc. Jpn.}\ }\textbf {\bibinfo {volume} {79}},\ \bibinfo {pages}
  {011003} (\bibinfo {year} {2010})}\BibitemShut {NoStop}%
\bibitem [{\citenamefont {Schmidt}\ and\ \citenamefont
  {Thalmeier}(2014)}]{schmidt:14}%
  \BibitemOpen
  \bibfield  {author} {\bibinfo {author} {\bibfnamefont {B.}~\bibnamefont
  {Schmidt}}\ and\ \bibinfo {author} {\bibfnamefont {P.}~\bibnamefont
  {Thalmeier}},\ }\href {\doibase 10.1103/PhysRevB.89.184402} {\bibfield
  {journal} {\bibinfo  {journal} {Phys. Rev. B}\ }\textbf {\bibinfo {volume}
  {89}},\ \bibinfo {pages} {184402} (\bibinfo {year} {2014})}\BibitemShut
  {NoStop}%
\bibitem [{\citenamefont {Kini}\ \emph {et~al.}(2006)\citenamefont {Kini},
  \citenamefont {Kaul},\ and\ \citenamefont {Geibel}}]{kini:06}%
  \BibitemOpen
  \bibfield  {author} {\bibinfo {author} {\bibfnamefont {N.~S.}\ \bibnamefont
  {Kini}}, \bibinfo {author} {\bibfnamefont {E.~E.}\ \bibnamefont {Kaul}}, \
  and\ \bibinfo {author} {\bibfnamefont {C.}~\bibnamefont {Geibel}},\ }\href
  {\doibase 10.1088/0953-8984/18/4/015} {\bibfield  {journal} {\bibinfo
  {journal} {J. Phys.: Cond. Mat.}\ }\textbf {\bibinfo {volume} {18}},\
  \bibinfo {pages} {1303} (\bibinfo {year} {2006})}\BibitemShut {NoStop}%
\bibitem [{\citenamefont {Shannon}\ \emph {et~al.}(2004)\citenamefont
  {Shannon}, \citenamefont {Schmidt}, \citenamefont {Penc},\ and\ \citenamefont
  {Thalmeier}}]{shannon:04}%
  \BibitemOpen
  \bibfield  {author} {\bibinfo {author} {\bibfnamefont {N.}~\bibnamefont
  {Shannon}}, \bibinfo {author} {\bibfnamefont {B.}~\bibnamefont {Schmidt}},
  \bibinfo {author} {\bibfnamefont {K.}~\bibnamefont {Penc}}, \ and\ \bibinfo
  {author} {\bibfnamefont {P.}~\bibnamefont {Thalmeier}},\ }\href {\doibase
  10.1140/epjb/e2004-00156-3} {\bibfield  {journal} {\bibinfo  {journal} {Eur.
  Phys. J. B}\ }\textbf {\bibinfo {volume} {38}},\ \bibinfo {pages} {599}
  (\bibinfo {year} {2004})}\BibitemShut {NoStop}%
\bibitem [{\citenamefont {Schmidt}\ \emph {et~al.}(2007)\citenamefont
  {Schmidt}, \citenamefont {Thalmeier},\ and\ \citenamefont
  {Shannon}}]{schmidt:07b}%
  \BibitemOpen
  \bibfield  {author} {\bibinfo {author} {\bibfnamefont {B.}~\bibnamefont
  {Schmidt}}, \bibinfo {author} {\bibfnamefont {P.}~\bibnamefont {Thalmeier}},
  \ and\ \bibinfo {author} {\bibfnamefont {N.}~\bibnamefont {Shannon}},\ }\href
  {\doibase 10.1103/PhysRevB.76.125113} {\bibfield  {journal} {\bibinfo
  {journal} {Phys. Rev. B}\ }\textbf {\bibinfo {volume} {76}},\ \bibinfo
  {pages} {125113} (\bibinfo {year} {2007})}\BibitemShut {NoStop}%
\bibitem [{\citenamefont {Schmidt}\ \emph {et~al.}(2010)\citenamefont
  {Schmidt}, \citenamefont {Siahatgar}, \citenamefont {Thalmeier},\ and\
  \citenamefont {Tsirlin}}]{schmidt:10}%
  \BibitemOpen
  \bibfield  {author} {\bibinfo {author} {\bibfnamefont {B.}~\bibnamefont
  {Schmidt}}, \bibinfo {author} {\bibfnamefont {M.}~\bibnamefont {Siahatgar}},
  \bibinfo {author} {\bibfnamefont {P.}~\bibnamefont {Thalmeier}}, \ and\
  \bibinfo {author} {\bibfnamefont {A.~A.}\ \bibnamefont {Tsirlin}},\ }\href
  {\doibase http://dx.doi.org/10.1088/1742-6596/200/2/022055} {\bibfield
  {journal} {\bibinfo  {journal} {J. Phys.: Conf. Ser.}\ }\textbf {\bibinfo
  {volume} {200}},\ \bibinfo {pages} {022055} (\bibinfo {year}
  {2010})}\BibitemShut {NoStop}%
\bibitem [{\citenamefont {Kawamura}\ and\ \citenamefont
  {Miyashita}(1984)}]{kawamura:84}%
  \BibitemOpen
  \bibfield  {author} {\bibinfo {author} {\bibfnamefont {H.}~\bibnamefont
  {Kawamura}}\ and\ \bibinfo {author} {\bibfnamefont {S.}~\bibnamefont
  {Miyashita}},\ }\href {\doibase 10.1143/JPSJ.53.4138} {\bibfield  {journal}
  {\bibinfo  {journal} {J. Phys. Soc. Jpn.}\ }\textbf {\bibinfo {volume}
  {53}},\ \bibinfo {pages} {4138} (\bibinfo {year} {1984})}\BibitemShut
  {NoStop}%
\bibitem [{\citenamefont {Suzuki}\ and\ \citenamefont
  {Matsubara}(1998)}]{suzuki:98}%
  \BibitemOpen
  \bibfield  {author} {\bibinfo {author} {\bibfnamefont {N.}~\bibnamefont
  {Suzuki}}\ and\ \bibinfo {author} {\bibfnamefont {F.}~\bibnamefont
  {Matsubara}},\ }\href {\doibase 10.1103/PhysRevB.58.5169} {\bibfield
  {journal} {\bibinfo  {journal} {Phys. Rev. B}\ }\textbf {\bibinfo {volume}
  {58}},\ \bibinfo {pages} {5169} (\bibinfo {year} {1998})}\BibitemShut
  {NoStop}%
\bibitem [{\citenamefont {Kulagin}\ \emph {et~al.}(2013)\citenamefont
  {Kulagin}, \citenamefont {Prokof'ev}, \citenamefont {Starykh}, \citenamefont
  {Svistunov},\ and\ \citenamefont {Varney}}]{kulagin:13}%
  \BibitemOpen
  \bibfield  {author} {\bibinfo {author} {\bibfnamefont {S.~A.}\ \bibnamefont
  {Kulagin}}, \bibinfo {author} {\bibfnamefont {N.}~\bibnamefont {Prokof'ev}},
  \bibinfo {author} {\bibfnamefont {O.~A.}\ \bibnamefont {Starykh}}, \bibinfo
  {author} {\bibfnamefont {B.}~\bibnamefont {Svistunov}}, \ and\ \bibinfo
  {author} {\bibfnamefont {C.~N.}\ \bibnamefont {Varney}},\ }\href {\doibase
  10.1103/PhysRevLett.110.070601} {\bibfield  {journal} {\bibinfo  {journal}
  {Phys. Rev. Lett.}\ }\textbf {\bibinfo {volume} {110}},\ \bibinfo {pages}
  {070601} (\bibinfo {year} {2013})}\BibitemShut {NoStop}%
\bibitem [{\citenamefont {Melchy}\ and\ \citenamefont
  {Zhitomirsky}(2009)}]{melchy:09}%
  \BibitemOpen
  \bibfield  {author} {\bibinfo {author} {\bibfnamefont {P.-E.}\ \bibnamefont
  {Melchy}}\ and\ \bibinfo {author} {\bibfnamefont {M.~E.}\ \bibnamefont
  {Zhitomirsky}},\ }\href {\doibase 10.1103/PhysRevB.80.064411} {\bibfield
  {journal} {\bibinfo  {journal} {Phys. Rev. B}\ }\textbf {\bibinfo {volume}
  {80}},\ \bibinfo {pages} {064411} (\bibinfo {year} {2009})}\BibitemShut
  {NoStop}%
\bibitem [{\citenamefont {Tamura}\ and\ \citenamefont
  {Tanaka}(2013)}]{tamura:13}%
  \BibitemOpen
  \bibfield  {author} {\bibinfo {author} {\bibfnamefont {R.}~\bibnamefont
  {Tamura}}\ and\ \bibinfo {author} {\bibfnamefont {S.}~\bibnamefont
  {Tanaka}},\ }\href {\doibase 10.1103/PhysRevE.88.052138} {\bibfield
  {journal} {\bibinfo  {journal} {Phys. Rev. E}\ }\textbf {\bibinfo {volume}
  {88}},\ \bibinfo {pages} {052138} (\bibinfo {year} {2013})}\BibitemShut
  {NoStop}%
\bibitem [{\citenamefont {Mezio}\ \emph {et~al.}(2012)\citenamefont {Mezio},
  \citenamefont {Manuel}, \citenamefont {Singh},\ and\ \citenamefont
  {Trumper}}]{mezio:12}%
  \BibitemOpen
  \bibfield  {author} {\bibinfo {author} {\bibfnamefont {A.}~\bibnamefont
  {Mezio}}, \bibinfo {author} {\bibfnamefont {L.~O.}\ \bibnamefont {Manuel}},
  \bibinfo {author} {\bibfnamefont {R.~R.~P.}\ \bibnamefont {Singh}}, \ and\
  \bibinfo {author} {\bibfnamefont {A.~E.}\ \bibnamefont {Trumper}},\ }\href
  {\doibase 10.1088/1367-2630/14/12/123033} {\bibfield  {journal} {\bibinfo
  {journal} {New J. Phys.}\ }\textbf {\bibinfo {volume} {14}},\ \bibinfo
  {pages} {123033} (\bibinfo {year} {2012})}\BibitemShut {NoStop}%
\bibitem [{\citenamefont {Zheng}\ \emph {et~al.}(2005)\citenamefont {Zheng},
  \citenamefont {Singh}, \citenamefont {McKenzie},\ and\ \citenamefont
  {Coldea}}]{zheng:05}%
  \BibitemOpen
  \bibfield  {author} {\bibinfo {author} {\bibfnamefont {W.}~\bibnamefont
  {Zheng}}, \bibinfo {author} {\bibfnamefont {R.~R.~P.}\ \bibnamefont {Singh}},
  \bibinfo {author} {\bibfnamefont {R.~H.}\ \bibnamefont {McKenzie}}, \ and\
  \bibinfo {author} {\bibfnamefont {R.}~\bibnamefont {Coldea}},\ }\href
  {\doibase 10.1103/PhysRevB.71.134422} {\bibfield  {journal} {\bibinfo
  {journal} {Phys. Rev. B}\ }\textbf {\bibinfo {volume} {71}},\ \bibinfo
  {pages} {134422} (\bibinfo {year} {2005})}\BibitemShut {NoStop}%
\bibitem [{\citenamefont {Zheng}\ \emph {et~al.}(1999)\citenamefont {Zheng},
  \citenamefont {McKenzie},\ and\ \citenamefont {Singh}}]{zheng:99}%
  \BibitemOpen
  \bibfield  {author} {\bibinfo {author} {\bibfnamefont {W.}~\bibnamefont
  {Zheng}}, \bibinfo {author} {\bibfnamefont {R.~H.}\ \bibnamefont {McKenzie}},
  \ and\ \bibinfo {author} {\bibfnamefont {R.~R.~P.}\ \bibnamefont {Singh}},\
  }\href {\doibase 10.1103/PhysRevB.59.14367} {\bibfield  {journal} {\bibinfo
  {journal} {Phys. Rev. B}\ }\textbf {\bibinfo {volume} {59}},\ \bibinfo
  {pages} {14367} (\bibinfo {year} {1999})}\BibitemShut {NoStop}%
\bibitem [{\citenamefont {Oitmaa}\ \emph {et~al.}(2006)\citenamefont {Oitmaa},
  \citenamefont {Hamer},\ and\ \citenamefont {Zheng}}]{oitmaa:06}%
  \BibitemOpen
  \bibfield  {author} {\bibinfo {author} {\bibfnamefont {J.}~\bibnamefont
  {Oitmaa}}, \bibinfo {author} {\bibfnamefont {C.}~\bibnamefont {Hamer}}, \
  and\ \bibinfo {author} {\bibfnamefont {W.}~\bibnamefont {Zheng}},\ }\href
  {\doibase 10.2277/0521842425} {\emph {\bibinfo {title} {Series Expansion
  Methods for Strongly Interacting Lattice Models}}}\ (\bibinfo  {publisher}
  {Cambridge University Press},\ \bibinfo {year} {2006})\BibitemShut {NoStop}%
\bibitem [{\citenamefont {Rosner}\ \emph {et~al.}(2003)\citenamefont {Rosner},
  \citenamefont {Singh}, \citenamefont {Zheng}, \citenamefont {Oitmaa},\ and\
  \citenamefont {Pickett}}]{rosner:03}%
  \BibitemOpen
  \bibfield  {author} {\bibinfo {author} {\bibfnamefont {H.}~\bibnamefont
  {Rosner}}, \bibinfo {author} {\bibfnamefont {R.~R.~P.}\ \bibnamefont
  {Singh}}, \bibinfo {author} {\bibfnamefont {W.~H.}\ \bibnamefont {Zheng}},
  \bibinfo {author} {\bibfnamefont {J.}~\bibnamefont {Oitmaa}}, \ and\ \bibinfo
  {author} {\bibfnamefont {W.~E.}\ \bibnamefont {Pickett}},\ }\href {\doibase
  10.1103/PhysRevB.67.014416} {\bibfield  {journal} {\bibinfo  {journal} {Phys.
  Rev. B}\ }\textbf {\bibinfo {volume} {67}},\ \bibinfo {pages} {014416}
  (\bibinfo {year} {2003})}\BibitemShut {NoStop}%
\bibitem [{\citenamefont {Nath}\ \emph {et~al.}(2008)\citenamefont {Nath},
  \citenamefont {Tsirlin}, \citenamefont {Rosner},\ and\ \citenamefont
  {Geibel}}]{nath:08}%
  \BibitemOpen
  \bibfield  {author} {\bibinfo {author} {\bibfnamefont {R.}~\bibnamefont
  {Nath}}, \bibinfo {author} {\bibfnamefont {A.~A.}\ \bibnamefont {Tsirlin}},
  \bibinfo {author} {\bibfnamefont {H.}~\bibnamefont {Rosner}}, \ and\ \bibinfo
  {author} {\bibfnamefont {C.}~\bibnamefont {Geibel}},\ }\href {\doibase
  10.1103/PhysRevB.78.064422} {\bibfield  {journal} {\bibinfo  {journal} {Phys.
  Rev. B}\ }\textbf {\bibinfo {volume} {78}},\ \bibinfo {pages} {064422}
  (\bibinfo {year} {2008})}\BibitemShut {NoStop}%
\bibitem [{\citenamefont {Schmidt}\ \emph
  {et~al.}(2011{\natexlab{a}})\citenamefont {Schmidt}, \citenamefont
  {Lohmann},\ and\ \citenamefont {Richter}}]{schmidthj:11}%
  \BibitemOpen
  \bibfield  {author} {\bibinfo {author} {\bibfnamefont {H.-J.}\ \bibnamefont
  {Schmidt}}, \bibinfo {author} {\bibfnamefont {A.}~\bibnamefont {Lohmann}}, \
  and\ \bibinfo {author} {\bibfnamefont {J.}~\bibnamefont {Richter}},\ }\href
  {\doibase 10.1103/PhysRevB.84.104443} {\bibfield  {journal} {\bibinfo
  {journal} {Phys. Rev. B}\ }\textbf {\bibinfo {volume} {84}},\ \bibinfo
  {pages} {104443} (\bibinfo {year} {2011}{\natexlab{a}})}\BibitemShut
  {NoStop}%
\bibitem [{\citenamefont {Jakli{\v c}}\ and\ \citenamefont {Prelov{\v
  s}ek}(2000)}]{jaklic:00}%
  \BibitemOpen
  \bibfield  {author} {\bibinfo {author} {\bibfnamefont {J.}~\bibnamefont
  {Jakli{\v c}}}\ and\ \bibinfo {author} {\bibfnamefont {P.}~\bibnamefont
  {Prelov{\v s}ek}},\ }\href {\doibase 10.1080/000187300243381} {\bibfield
  {journal} {\bibinfo  {journal} {Adv. Phys.}\ }\textbf {\bibinfo {volume}
  {49}},\ \bibinfo {pages} {1} (\bibinfo {year} {2000})}\BibitemShut {NoStop}%
\bibitem [{\citenamefont {Misguich}\ \emph {et~al.}(2003)\citenamefont
  {Misguich}, \citenamefont {Bernu},\ and\ \citenamefont
  {Pierre}}]{misguich:03}%
  \BibitemOpen
  \bibfield  {author} {\bibinfo {author} {\bibfnamefont {G.}~\bibnamefont
  {Misguich}}, \bibinfo {author} {\bibfnamefont {B.}~\bibnamefont {Bernu}}, \
  and\ \bibinfo {author} {\bibfnamefont {L.}~\bibnamefont {Pierre}},\ }\href
  {\doibase 10.1103/PhysRevB.68.113409} {\bibfield  {journal} {\bibinfo
  {journal} {Phys. Rev. B}\ }\textbf {\bibinfo {volume} {68}},\ \bibinfo
  {pages} {113409} (\bibinfo {year} {2003})}\BibitemShut {NoStop}%
\bibitem [{\citenamefont {Schmidt}\ \emph {et~al.}(2009)\citenamefont
  {Schmidt}, \citenamefont {Thalmeier},\ and\ \citenamefont
  {Shannon}}]{schmidt:08}%
  \BibitemOpen
  \bibfield  {author} {\bibinfo {author} {\bibfnamefont {B.}~\bibnamefont
  {Schmidt}}, \bibinfo {author} {\bibfnamefont {P.}~\bibnamefont {Thalmeier}},
  \ and\ \bibinfo {author} {\bibfnamefont {N.}~\bibnamefont {Shannon}},\ }\href
  {\doibase 10.1088/1742-6596/145/1/012054} {\bibfield  {journal} {\bibinfo
  {journal} {J. Phys.: Conf. Ser.}\ }\textbf {\bibinfo {volume} {145}},\
  \bibinfo {pages} {012054} (\bibinfo {year} {2009})}\BibitemShut {NoStop}%
\bibitem [{\citenamefont {Hanebaum}\ and\ \citenamefont
  {Schnack}(2014)}]{hanebaum:14}%
  \BibitemOpen
  \bibfield  {author} {\bibinfo {author} {\bibfnamefont {O.}~\bibnamefont
  {Hanebaum}}\ and\ \bibinfo {author} {\bibfnamefont {J.}~\bibnamefont
  {Schnack}},\ }\href {\doibase 10.1140/epjb/e2014-50360-5} {\bibfield
  {journal} {\bibinfo  {journal} {Eur. Phys. J. B}\ }\textbf {\bibinfo {volume}
  {87}},\ \bibinfo {eid} {194} (\bibinfo {year} {2014})}\BibitemShut {NoStop}%
\bibitem [{\citenamefont {Lanczos}(1950)}]{lanczos:50}%
  \BibitemOpen
  \bibfield  {author} {\bibinfo {author} {\bibfnamefont {C.}~\bibnamefont
  {Lanczos}},\ }\href {\doibase 10.6028/jres.045.026} {\bibfield  {journal}
  {\bibinfo  {journal} {J. Res. Natl. Bur. Stand.}\ }\textbf {\bibinfo {volume}
  {45}},\ \bibinfo {pages} {255} (\bibinfo {year} {1950})}\BibitemShut
  {NoStop}%
\bibitem [{\citenamefont {Schmidt}\ \emph
  {et~al.}(2011{\natexlab{b}})\citenamefont {Schmidt}, \citenamefont
  {Siahatgar},\ and\ \citenamefont {Thalmeier}}]{schmidt:11}%
  \BibitemOpen
  \bibfield  {author} {\bibinfo {author} {\bibfnamefont {B.}~\bibnamefont
  {Schmidt}}, \bibinfo {author} {\bibfnamefont {M.}~\bibnamefont {Siahatgar}},
  \ and\ \bibinfo {author} {\bibfnamefont {P.}~\bibnamefont {Thalmeier}},\
  }\href {\doibase 10.1103/PhysRevB.83.075123} {\bibfield  {journal} {\bibinfo
  {journal} {Phys. Rev. B}\ }\textbf {\bibinfo {volume} {83}},\ \bibinfo
  {pages} {075123} (\bibinfo {year} {2011}{\natexlab{b}})}\BibitemShut
  {NoStop}%
\bibitem [{\citenamefont {Thesberg}\ and\ \citenamefont
  {S\o{}rensen}(2014)}]{thesberg:14}%
  \BibitemOpen
  \bibfield  {author} {\bibinfo {author} {\bibfnamefont {M.}~\bibnamefont
  {Thesberg}}\ and\ \bibinfo {author} {\bibfnamefont {E.~S.}\ \bibnamefont
  {S\o{}rensen}},\ }\href {\doibase 10.1103/PhysRevB.90.115117} {\bibfield
  {journal} {\bibinfo  {journal} {Phys. Rev. B}\ }\textbf {\bibinfo {volume}
  {90}},\ \bibinfo {pages} {115117} (\bibinfo {year} {2014})}\BibitemShut
  {NoStop}%
\bibitem [{\citenamefont {Bonner}\ and\ \citenamefont
  {Fisher}(1964)}]{bonner:64}%
  \BibitemOpen
  \bibfield  {author} {\bibinfo {author} {\bibfnamefont {J.~C.}\ \bibnamefont
  {Bonner}}\ and\ \bibinfo {author} {\bibfnamefont {M.~E.}\ \bibnamefont
  {Fisher}},\ }\href {\doibase 10.1103/PhysRev.135.A640} {\bibfield  {journal}
  {\bibinfo  {journal} {Phys. Rev.}\ }\textbf {\bibinfo {volume} {135}},\
  \bibinfo {pages} {A640} (\bibinfo {year} {1964})}\BibitemShut {NoStop}%
\bibitem [{\citenamefont {Cong}\ \emph {et~al.}(2011)\citenamefont {Cong},
  \citenamefont {Wolf}, \citenamefont {de~Souza}, \citenamefont {Kr\"uger},
  \citenamefont {Haghighirad}, \citenamefont {Gottlieb-Schoenmeyer},
  \citenamefont {Ritter}, \citenamefont {Assmus}, \citenamefont {Opahle},
  \citenamefont {Foyevtsova}, \citenamefont {Jeschke}, \citenamefont
  {Valent\'\i}, \citenamefont {Wiehl},\ and\ \citenamefont {Lang}}]{cong:11}%
  \BibitemOpen
  \bibfield  {author} {\bibinfo {author} {\bibfnamefont {P.~T.}\ \bibnamefont
  {Cong}}, \bibinfo {author} {\bibfnamefont {B.}~\bibnamefont {Wolf}}, \bibinfo
  {author} {\bibfnamefont {M.}~\bibnamefont {de~Souza}}, \bibinfo {author}
  {\bibfnamefont {N.}~\bibnamefont {Kr\"uger}}, \bibinfo {author}
  {\bibfnamefont {A.~A.}\ \bibnamefont {Haghighirad}}, \bibinfo {author}
  {\bibfnamefont {S.}~\bibnamefont {Gottlieb-Schoenmeyer}}, \bibinfo {author}
  {\bibfnamefont {F.}~\bibnamefont {Ritter}}, \bibinfo {author} {\bibfnamefont
  {W.}~\bibnamefont {Assmus}}, \bibinfo {author} {\bibfnamefont
  {I.}~\bibnamefont {Opahle}}, \bibinfo {author} {\bibfnamefont
  {K.}~\bibnamefont {Foyevtsova}}, \bibinfo {author} {\bibfnamefont {H.~O.}\
  \bibnamefont {Jeschke}}, \bibinfo {author} {\bibfnamefont {R.}~\bibnamefont
  {Valent\'\i}}, \bibinfo {author} {\bibfnamefont {L.}~\bibnamefont {Wiehl}}, \
  and\ \bibinfo {author} {\bibfnamefont {M.}~\bibnamefont {Lang}},\ }\href
  {\doibase 10.1103/PhysRevB.83.064425} {\bibfield  {journal} {\bibinfo
  {journal} {Phys. Rev. B}\ }\textbf {\bibinfo {volume} {83}},\ \bibinfo
  {pages} {064425} (\bibinfo {year} {2011})}\BibitemShut {NoStop}%
\bibitem [{\citenamefont {Coldea}\ \emph {et~al.}(2002)\citenamefont {Coldea},
  \citenamefont {Tennant}, \citenamefont {Habicht}, \citenamefont {Smeibidl},
  \citenamefont {Wolters},\ and\ \citenamefont {Tylczynski}}]{coldea:02}%
  \BibitemOpen
  \bibfield  {author} {\bibinfo {author} {\bibfnamefont {R.}~\bibnamefont
  {Coldea}}, \bibinfo {author} {\bibfnamefont {D.~A.}\ \bibnamefont {Tennant}},
  \bibinfo {author} {\bibfnamefont {K.}~\bibnamefont {Habicht}}, \bibinfo
  {author} {\bibfnamefont {P.}~\bibnamefont {Smeibidl}}, \bibinfo {author}
  {\bibfnamefont {C.}~\bibnamefont {Wolters}}, \ and\ \bibinfo {author}
  {\bibfnamefont {Z.}~\bibnamefont {Tylczynski}},\ }\href {\doibase
  10.1103/PhysRevLett.88.137203} {\bibfield  {journal} {\bibinfo  {journal}
  {Phys. Rev. Lett.}\ }\textbf {\bibinfo {volume} {88}},\ \bibinfo {pages}
  {137203} (\bibinfo {year} {2002})}\BibitemShut {NoStop}%
\bibitem [{\citenamefont {Zvyagin}\ \emph {et~al.}(2014)\citenamefont
  {Zvyagin}, \citenamefont {Kamenskyi}, \citenamefont {Ozerov}, \citenamefont
  {Wosnitza}, \citenamefont {Ikeda}, \citenamefont {Fujita}, \citenamefont
  {Hagiwara}, \citenamefont {Smirnov}, \citenamefont {Soldatov}, \citenamefont
  {Shapiro}, \citenamefont {Krzystek}, \citenamefont {Hu}, \citenamefont {Ryu},
  \citenamefont {Petrovic},\ and\ \citenamefont {Zhitomirsky}}]{zvyagin:14}%
  \BibitemOpen
  \bibfield  {author} {\bibinfo {author} {\bibfnamefont {S.~A.}\ \bibnamefont
  {Zvyagin}}, \bibinfo {author} {\bibfnamefont {D.}~\bibnamefont {Kamenskyi}},
  \bibinfo {author} {\bibfnamefont {M.}~\bibnamefont {Ozerov}}, \bibinfo
  {author} {\bibfnamefont {J.}~\bibnamefont {Wosnitza}}, \bibinfo {author}
  {\bibfnamefont {M.}~\bibnamefont {Ikeda}}, \bibinfo {author} {\bibfnamefont
  {T.}~\bibnamefont {Fujita}}, \bibinfo {author} {\bibfnamefont
  {M.}~\bibnamefont {Hagiwara}}, \bibinfo {author} {\bibfnamefont
  {A.}~\bibnamefont {Smirnov}}, \bibinfo {author} {\bibfnamefont
  {T.}~\bibnamefont {Soldatov}}, \bibinfo {author} {\bibfnamefont
  {A.}~\bibnamefont {Shapiro}}, \bibinfo {author} {\bibfnamefont
  {J.}~\bibnamefont {Krzystek}}, \bibinfo {author} {\bibfnamefont
  {R.}~\bibnamefont {Hu}}, \bibinfo {author} {\bibfnamefont {H.}~\bibnamefont
  {Ryu}}, \bibinfo {author} {\bibfnamefont {C.}~\bibnamefont {Petrovic}}, \
  and\ \bibinfo {author} {\bibfnamefont {M.}~\bibnamefont {Zhitomirsky}},\
  }\href {\doibase 10.1103/PhysRevLett.112.077206} {\bibfield  {journal}
  {\bibinfo  {journal} {Phys. Rev. Lett.}\ }\textbf {\bibinfo {volume} {112}},\
  \bibinfo {pages} {077206} (\bibinfo {year} {2014})}\BibitemShut {NoStop}%
\bibitem [{\citenamefont {Carlin}\ \emph {et~al.}(1985)\citenamefont {Carlin},
  \citenamefont {Burriel}, \citenamefont {Palacio}, \citenamefont {Carlin},
  \citenamefont {Keij},\ and\ \citenamefont {Carnegie}}]{carlin:85}%
  \BibitemOpen
  \bibfield  {author} {\bibinfo {author} {\bibfnamefont {R.~L.}\ \bibnamefont
  {Carlin}}, \bibinfo {author} {\bibfnamefont {R.}~\bibnamefont {Burriel}},
  \bibinfo {author} {\bibfnamefont {F.}~\bibnamefont {Palacio}}, \bibinfo
  {author} {\bibfnamefont {R.~A.}\ \bibnamefont {Carlin}}, \bibinfo {author}
  {\bibfnamefont {S.~F.}\ \bibnamefont {Keij}}, \ and\ \bibinfo {author}
  {\bibfnamefont {D.~W.}\ \bibnamefont {Carnegie}},\ }\href {\doibase
  10.1063/1.335093} {\bibfield  {journal} {\bibinfo  {journal} {J. Appl.
  Phys.}\ }\textbf {\bibinfo {volume} {57}},\ \bibinfo {pages} {3351} (\bibinfo
  {year} {1985})}\BibitemShut {NoStop}%
\bibitem [{\citenamefont {Coldea}\ \emph {et~al.}(2003)\citenamefont {Coldea},
  \citenamefont {Tennant},\ and\ \citenamefont {Tylczynski}}]{coldea:03}%
  \BibitemOpen
  \bibfield  {author} {\bibinfo {author} {\bibfnamefont {R.}~\bibnamefont
  {Coldea}}, \bibinfo {author} {\bibfnamefont {D.~A.}\ \bibnamefont {Tennant}},
  \ and\ \bibinfo {author} {\bibfnamefont {Z.}~\bibnamefont {Tylczynski}},\
  }\href {\doibase 10.1103/PhysRevB.68.134424} {\bibfield  {journal} {\bibinfo
  {journal} {Phys. Rev. B}\ }\textbf {\bibinfo {volume} {68}},\ \bibinfo
  {pages} {134424} (\bibinfo {year} {2003})}\BibitemShut {NoStop}%
\bibitem [{\citenamefont {Cong}\ \emph {et~al.}(2014)\citenamefont {Cong},
  \citenamefont {Wolf}, \citenamefont {van Well}, \citenamefont {Haghighirad},
  \citenamefont {Ritter}, \citenamefont {Assmus}, \citenamefont {Krellner},\
  and\ \citenamefont {Lang}}]{cong:14}%
  \BibitemOpen
  \bibfield  {author} {\bibinfo {author} {\bibfnamefont {P.~T.}\ \bibnamefont
  {Cong}}, \bibinfo {author} {\bibfnamefont {B.}~\bibnamefont {Wolf}}, \bibinfo
  {author} {\bibfnamefont {N.}~\bibnamefont {van Well}}, \bibinfo {author}
  {\bibfnamefont {A.}~\bibnamefont {Haghighirad}}, \bibinfo {author}
  {\bibfnamefont {F.}~\bibnamefont {Ritter}}, \bibinfo {author} {\bibfnamefont
  {W.}~\bibnamefont {Assmus}}, \bibinfo {author} {\bibfnamefont
  {C.}~\bibnamefont {Krellner}}, \ and\ \bibinfo {author} {\bibfnamefont
  {M.}~\bibnamefont {Lang}},\ }\href {\doibase 10.1109/TMAG.2014.2298496}
  {\bibfield  {journal} {\bibinfo  {journal} {IEEE Transactions on Magnetics}\
  }\textbf {\bibinfo {volume} {50}},\ \bibinfo {pages} {1} (\bibinfo {year}
  {2014})}\BibitemShut {NoStop}%
\bibitem [{sch()}]{schmidt:15b}%
  \BibitemOpen
  \href@noop {} {}\bibinfo {note} {Unpublished}\BibitemShut {NoStop}%
\bibitem [{\citenamefont {Itou}\ \emph {et~al.}(2008)\citenamefont {Itou},
  \citenamefont {Oyamada}, \citenamefont {Maegawa}, \citenamefont {Tamura},\
  and\ \citenamefont {Kato}}]{itou:08}%
  \BibitemOpen
  \bibfield  {author} {\bibinfo {author} {\bibfnamefont {T.}~\bibnamefont
  {Itou}}, \bibinfo {author} {\bibfnamefont {A.}~\bibnamefont {Oyamada}},
  \bibinfo {author} {\bibfnamefont {S.}~\bibnamefont {Maegawa}}, \bibinfo
  {author} {\bibfnamefont {M.}~\bibnamefont {Tamura}}, \ and\ \bibinfo {author}
  {\bibfnamefont {R.}~\bibnamefont {Kato}},\ }\href {\doibase
  10.1103/PhysRevB.77.104413} {\bibfield  {journal} {\bibinfo  {journal} {Phys.
  Rev. B}\ }\textbf {\bibinfo {volume} {77}},\ \bibinfo {pages} {104413}
  (\bibinfo {year} {2008})}\BibitemShut {NoStop}%
\bibitem [{\citenamefont {Tamura}\ \emph {et~al.}(2006)\citenamefont {Tamura},
  \citenamefont {Nakao},\ and\ \citenamefont {Kato}}]{tamura:06}%
  \BibitemOpen
  \bibfield  {author} {\bibinfo {author} {\bibfnamefont {M.}~\bibnamefont
  {Tamura}}, \bibinfo {author} {\bibfnamefont {A.}~\bibnamefont {Nakao}}, \
  and\ \bibinfo {author} {\bibfnamefont {R.}~\bibnamefont {Kato}},\ }\href
  {\doibase 10.1143/JPSJ.75.093701} {\bibfield  {journal} {\bibinfo  {journal}
  {J. Phys. Soc. Jpn.}\ }\textbf {\bibinfo {volume} {75}},\ \bibinfo {pages}
  {093701} (\bibinfo {year} {2006})}\BibitemShut {NoStop}%
\bibitem [{\citenamefont {Yoshida}\ \emph {et~al.}(2015)\citenamefont
  {Yoshida}, \citenamefont {Ito}, \citenamefont {Maesato}, \citenamefont
  {Shimizu}, \citenamefont {Hayama}, \citenamefont {Hiramatsu}, \citenamefont
  {Nakamura}, \citenamefont {Kishida}, \citenamefont {Koretsune}, \citenamefont
  {Hotta},\ and\ \citenamefont {Saito}}]{yoshida:15}%
  \BibitemOpen
  \bibfield  {author} {\bibinfo {author} {\bibfnamefont {Y.}~\bibnamefont
  {Yoshida}}, \bibinfo {author} {\bibfnamefont {H.}~\bibnamefont {Ito}},
  \bibinfo {author} {\bibfnamefont {M.}~\bibnamefont {Maesato}}, \bibinfo
  {author} {\bibfnamefont {Y.}~\bibnamefont {Shimizu}}, \bibinfo {author}
  {\bibfnamefont {H.}~\bibnamefont {Hayama}}, \bibinfo {author} {\bibfnamefont
  {T.}~\bibnamefont {Hiramatsu}}, \bibinfo {author} {\bibfnamefont
  {Y.}~\bibnamefont {Nakamura}}, \bibinfo {author} {\bibfnamefont
  {H.}~\bibnamefont {Kishida}}, \bibinfo {author} {\bibfnamefont
  {T.}~\bibnamefont {Koretsune}}, \bibinfo {author} {\bibfnamefont
  {C.}~\bibnamefont {Hotta}}, \ and\ \bibinfo {author} {\bibfnamefont
  {G.}~\bibnamefont {Saito}},\ }\href {\doibase 10.1038/nphys3359} {\bibfield
  {journal} {\bibinfo  {journal} {Nat Phys}\ }\textbf {\bibinfo {volume}
  {advance online publication}},\  (\bibinfo {year} {2015})}\BibitemShut
  {NoStop}%
\bibitem [{\citenamefont {Hamad}\ \emph {et~al.}(2005)\citenamefont {Hamad},
  \citenamefont {Trumper}, \citenamefont {Wzietek}, \citenamefont {Lefebvre},\
  and\ \citenamefont {Manuel}}]{hamad:05}%
  \BibitemOpen
  \bibfield  {author} {\bibinfo {author} {\bibfnamefont {I.~J.}\ \bibnamefont
  {Hamad}}, \bibinfo {author} {\bibfnamefont {A.~E.}\ \bibnamefont {Trumper}},
  \bibinfo {author} {\bibfnamefont {P.}~\bibnamefont {Wzietek}}, \bibinfo
  {author} {\bibfnamefont {S.}~\bibnamefont {Lefebvre}}, \ and\ \bibinfo
  {author} {\bibfnamefont {L.~O.}\ \bibnamefont {Manuel}},\ }\href {\doibase
  10.1088/0953-8984/17/50/026} {\bibfield  {journal} {\bibinfo  {journal} {J.
  Phys.: Cond. Mat.}\ }\textbf {\bibinfo {volume} {17}},\ \bibinfo {pages}
  {8091} (\bibinfo {year} {2005})}\BibitemShut {NoStop}%
\end{thebibliography}%

\end{document}